\documentclass[journal,onecolumn,12pt,twoside]{IEEEtran}

\usepackage{subfigure}
\usepackage{bm}
\usepackage{times}
\usepackage{epsf,epsfig}
\usepackage{float}
\usepackage{placeins}
\usepackage{psfrag}
\usepackage{amsmath}
\usepackage{bm}
\usepackage{amssymb}
\usepackage{graphicx}
\usepackage{epstopdf}
\usepackage{multirow}
\usepackage{enumerate}
\usepackage{color}
\usepackage{mathrsfs}
\usepackage{setspace}
\usepackage{diagbox}
\usepackage{enumitem}

\makeatletter
\newif\if@restonecol
\makeatother

\usepackage[linesnumbered,ruled,vlined]{algorithm2e}
\usepackage{algpseudocode}

\ifCLASSINFOpdf
\else
\fi
\begin{document}
%

\title{Hardware Architecture of Layered Decoders for PLDPC-Hadamard Codes}

\author{
Peng W.  Zhang, 
Sheng  Jiang,
        Francis C.M. Lau,~\textit{Fellow,~IEEE,} \\
        and~Chiu-W. Sham,~\textit{Senior Member,~IEEE}
\thanks{P.~W. Zhang was  with the Future Wireless Networks and IoT Focusing Area,
         Department
of Electronic and Information Engineering, The Hong Kong Polytechnic University, Hong Kong SAR, China. He is now with Huawei Technologies Ltd., Chengdu, China  (e-mail: pengwei.zhang@connect.polyu.hk).}
      \thanks{ S. Jiang and  F.~C.~M. Lau are with the Future Wireless Networks and IoT Focusing Area,
         Department
of Electronic and Information Engineering, The Hong Kong Polytechnic University, Hong Kong SAR, China  (e-mail: sheng.jiang@connect.polyu.hk; francis-cm.lau@polyu.edu.hk).}
\thanks{C.-W. Sham is with the Department of Computer Science,  The University of Auckland,
New Zealand (e-mail: b.sham@auckland.ac.nz).}
\thanks{ The work described in this paper was partially supported by the Postdoc Matching Fund Scheme, The Hong Kong Polytechnic University, Hong Kong SAR, China (Project ID P0035802).}
}
\maketitle

\begin{abstract}
Protograph-based low-density parity-check Hadamard codes (PLDPC-HCs)
are a new type of ultimate-Shannon-limit-approaching codes.
In this paper, we propose a hardware architecture for the  PLDPC-HC layered decoders.
The decoders consist mainly of random address memories, Hadamard sub-decoders and control logics.
Two types of pipelined structures are presented and the latency and throughput of these two structures are
derived.
Implementation of the decoder design on an FPGA board shows that
a throughput of $1.48$ Gbps is achieved with a bit error rate (BER) of $10^{-5}$
 at around $E_b/N_0=-0.40$ dB.
 The decoder can also achieve the same BER at {\color{black}$E_b/N_0=-1.14$ dB} with
 a reduced throughput of $0.20$ Gbps.
%
\end{abstract}


%
\IEEEpeerreviewmaketitle


\section{Introduction}\label{sect:Introduction}
Both turbo codes \cite{Berrou1993} and low-density parity-check (LDPC) codes \cite{Gallager1963}
have been demonstrated to be capacity-approaching channel codes \cite{Divsalar1998,Richardson2001}.
They have been used in a wide variety of communication
and data storage systems
\cite{Brejza2016}, including 3G/4G/5G
cellular communications,
optical communications, and magnetic recording systems \cite{Ardakani2015,Liu2018,Fang2018};
 and
various encoder/decoder designs have been proposed
\cite{Li2006,Cheng2014,Zhao2015,Lee2016,Hailes2016}.
Among different types of the LDPC codes, the structured quasi-cyclic (QC) LDPC codes allow easy realization
of linear encoding and parallel decoding.
QC-LDPC codes can be constructed from the perspective of a protograph.
 By lifting a protograph containing a small number of variable nodes and check nodes,
QC-LDPC codes called protograph-based LDPC (PLDPC) codes are formed \cite{Thorpe2003, Fang2015}.
It has also been shown that well-designed QC-LDPC codes
can achieve good decoding performance, low error floor and high throughput
%

{\color{black} To decode QC-LDPC codes, layered decoding architectures
are mostly used because they have relatively low hardware requirements and high throughputs. 
 For example, a rate-compatible layered decoding architecture that
 allows  parallel decoding of QC-LDPC codes has been shown to achieve a throughput of $1.28$ Gbps \cite{ZhangK2011}.
 In \cite{Marchand2009}, it has been shown that memory access conflicts introduced by the pipeline process in layered decoding can be reduced by 
 lowering the maximum available parallelism and efficient scheduling. 
 %
In \cite{Sun2013}, a novel layered decoder architecture that supports
QC-LDPC codes with any circulant weight is proposed. 
To resolve the access conflict issue, a block-serial scheduling algorithm, whose processing time is independent of the circulant weight, is further developed. 
Using the China Mobile Multimedia Broadcasting standard as an example, a decoder  synthesized using $65$-nm CMOS technology has shown to achieve a throughput of $1.1$~Gb/s with $15$ iterations.
In \cite{Kumawat2015}, a block-level-parallel
layered decoder for irregular QC-LDPC codes is proposed and
a dynamic multi-frame processing schedule is developed to minimize pipeline stages and memory overheads. The decoder can also be 
 reconfigured to support multiple block lengths and code rates of the WiFi standard. 
In \cite{Lu2016}, a RAM-based decoder architecture is proposed to decode cyclically-coupled QC-LDPC codes
and obtains a throughput of $3.0$ Gbps and an error floor of about $10^{-16}$.
In  \cite{Lee2017}, it is shown that a layered decoder throughput 
can be increased 
by reordering the layered decoding procedure and applying some optimization techniques.
In \cite{Boncalo2019}, off-line mapping and scheduling algorithms
have been proposed together with 
a novel residue-based layered QC-LDPC decoding 
to increase the resource usage of the layered decoder.   
Evaluation performed for six
QC-LDPC codes shows up to 57\% improvement 
in hardware utilization efficiency for a one-layer
overlap. 
In \cite{Petrovic2020}, an efficient decoder architecture is proposed for 
highly irregular QC-LDPC codes. 
It normally works as the layered schedule.
When a pipeline conflict is foreseen, the decoder changes to the flooding
schedule.
An offline parity-check matrix reordering method based on 
genetic algorithm is then further to optimized this hybrid schedule. 
Throughput
increases between 30.8\% and 109.1\% are demonstrated for 5G NR codes.
In  \cite{Li2021}, it is shown that 
with a multi-core architecture and a full row-parallel layered decoder, a throughput of $860$ Gbps is achievable
at a maximum of $2$ decoding iterations. 
%
%
In \cite{Verma2021}, a logarithmic-likelihood-ratio compound (LLRC) segregation technique is proposed. Based on the technique and 
other architectural optimizations, a 
hardware-efficient QC-LDPC layered decoder architecture with
reduced data-congestion and high throughput is presented. 
Comparison with other
works shows that the proposed decoder achieves more than two times  throughput improvement 
and eight times better hardware-efficiency.
}

Moreover, when both turbo and LDPC codes are used together with Hadamard codes, forming turbo-Hadamard codes
 \cite{Li2003} and LDPC-Hadamard codes (LDPC-HCs) \cite{Yue2007}, respectively,
 very good error performance can be achieved even when operating 
close to the ultimate Shannon limit (i.e., bit-energy-to-noise-power-spectral-density ratio ($E_b /N_0$) equals $-1.59$ dB) \cite{Costello2007}.
Another ultimate-Shannon-limit-approaching code is the
concatenated zigzag-Hadamard code \cite{Leung2006}.
Among these three types of codes, LDPC-HCs have been shown to produce
the best error performance.
For example, a rate-$0.05$ LDPC-HC with a theoretical threshold of $-1.35$ dB can achieve a bit error rate (BER) of $10^{-5}$ at $E_b /N_0  = -1.18$ dB \cite{Yue2007}.
These ultimate-Shannon-limit-approaching codes can be applied to extreme communication
environments such as deep-space communications and interleave division multiple access
systems with many users \cite{LiPing2006}.

Recently, a new type of LDPC-HCs called
protograph-based LDPC Hadamard codes (PLDPC-HCs) have been proposed, and
a new technique is developed to
enable the analysis of PLDPC-HCs which possess degree-$1$ and/or punctured variable nodes
\cite{zhang2020proto,zhang2021}. 
PLDPC-HCs perform as good as traditional LDPC-HCs.
For instance, a rate-$0.0494$ PLDPC-HC with a theoretical threshold of $-1.42$ dB
is found to achieve a BER of $10^{-5}$ at $E_b /N_0  = -1.19$ dB.
In addition, PLDPC-HC possesses a {\color{black} semi-regular \footnote{In the protograph of a PLDPC-HC \cite{zhang2020proto,zhang2021}, the degrees of the protograph variable nodes can be different
 while the degrees of Hadamard check nodes are kept the same.}} quasi-cyclic structure which is beneficial to hardware implementation.
To improve the convergence rate, a  PLDPC-HC layered decoding algorithm
has been proposed \cite{Zhang2021layer}. In this paper,
we propose a hardware architecture for PLDPC-HC layered decoders.
The proposed architecture is generic and can be readily modified to decode
other PLDPC-derived codes when the Hadamard constraint in the PLDPC-HC
 is replaced by other coding constraints.

The  paper is organized as follows.
Section \ref{sect:str_LDPCH} reviews the structure of a PLDPC-HC and its layered decoding algorithm.
Section \ref{sect:hw_LDPCH} first introduces the read and write operations of a
random access memory and the pipeline structure of a Hadamard sub-decoder.
Then it presents a hardware architecture of PLDPC-HC layered decoders, and derives its
 latency and throughput.
Section \ref{sect:results} shows the implementation results and
finally Section \ref{sect:conclusion} gives some concluding remarks.

\begin{figure*}[t]
\centering
\begin{minipage}[c]{0.6\textwidth}
    \begin{equation}\label{mat:r=4}
    {\bm{B}_{7 \times 11}} = \left[ {\begin{array}{*{11}{c}}
    1&0&0&0&0&0&1&0&3&0&1\\
    0&1&2&0&0&0&0&0&0&2&1\\
    2&1&0&0&1&1&0&0&0&0&1\\
    0&1&0&3&0&0&0&0&0&2&0\\
    2&0&0&0&0&0&0&1&0&3&0\\
    3&0&0&2&0&0&1&0&0&0&0\\
    1&0&0&1&1&0&0&0&1&2&0
    \end{array}} \right]\nonumber
    \end{equation}
\end{minipage}%
\\

\begin{minipage}[c]{\textwidth}
\centering
    \includegraphics[width=.95\textwidth]{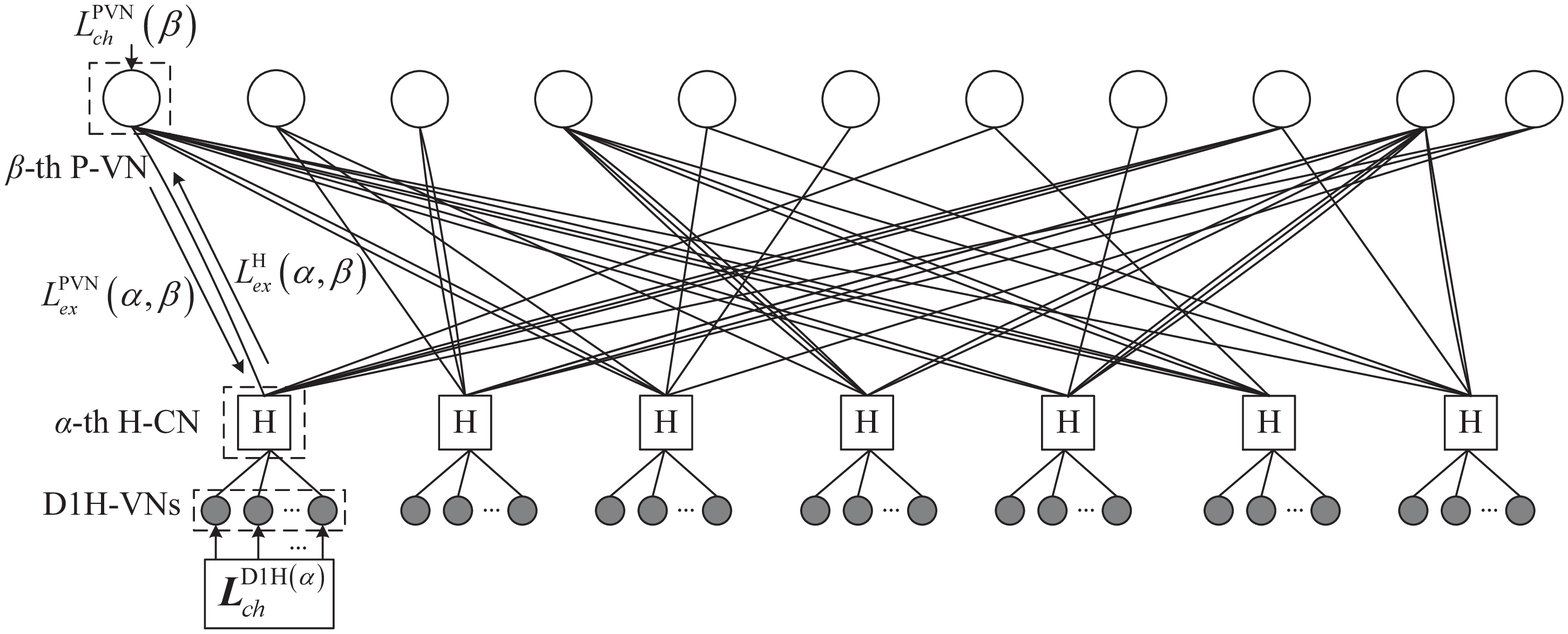}
\end{minipage}
\caption{The base matrix and corresponding protograph of a PLDPC-Hadamard code  \cite{zhang2020proto,zhang2021}.
A circle denotes a protograph variable node (P-VN), a square with $\rm ``H"$ denotes a Hadamard check node (H-CN), and a filled circle denotes a degree-1 Hadamard variable node (D1H-VN).
Row weight $d=6$, Hadamard order $r=d-2=4$, and $2^r-r-2=10$ D1H-VNs are attached to each H-CN.
Code rate $R = 0.0494$.}
\label{fig:base&proto}
\end{figure*}

\section{Review of PLDPC-Hadamard Codes}\label{sect:str_LDPCH}
The structure of a PLDPC-HC can be constructed from a PLDPC code \cite{Thorpe2003}.
When the check nodes in a PLDPC code are replaced by Hadamard check-nodes (H-CNs)
to which an appropriate number of degree-$1$ Hadamard variable nodes (D1H-VNs)
are connected, a PLDPC-HC is formed \cite{zhang2020proto,zhang2021}.
Fig.~\ref{fig:base&proto} illustrates the base matrix $\bm{B}_{m \times n}$
of a PLDPC-HC and its corresponding protograph.
As can be observed, there are $n=11$ protograph variable nodes (P-VNs) and $m=7$ H-CNs.
Moreover,  each H-CN is connected to a number of D1H-VNs.
The $(i,j)$-th entry in $\bm{B}_{m \times n}$, represented by $B(i,j)$, denotes the number of edges connected
between the $i$-th H-CN and the $j$-th P-VN.
In this example, each H-CN is connected to $d=6$ P-VNs,
where $d$ also equals the row weight of the base matrix $\bm{B}_{m \times n}$.
To obtain the adjacency matrix $\bm{H}_{M \times N}$  of the PLDPC-HC,
the base matrix $\bm{B}_{m \times n}$ is lifted twice with factors $z_1$ and $z_2$
where $M=m z_1 z_2$ and $N=n z_1 z_2$  \cite{Wang2013}.
The first lifting replaces each non-zero $B(i,j)$ in $\bm{B}_{m \times n}$ with
a summation of $B(i,j)$ different $z_1 \times z_1$  permutation matrices, and each $B(i,j) = 0$ with $z_1 \times z_1$ zero matrix.
The aim is to remove  parallel edges between P-VNs and H-CNs.
The second lifting then replaces each ``1'' with a circulant permutation matrix (CPM) of size
$z_2 \times z_2$ and each ``0'' with the $z_2 \times z_2$ zero matrix. The aim is to construct a quasi-cyclic code structure for easy encoding and decoding \cite{Fossorier2004}.
After the double-lifting process, the lifted graph, which corresponds to the adjacency matrix, contains $M$ H-CNs and $N$ P-VNs.

Based on the adjacency matrix $\bm{H}_{M \times N}$ obtained, $N-M$ information
bits are first encoded into a length-$N$ LDPC code. Then for each H-CN, the $d$ incoming
messages from the P-VNs are used to encode an order $r$ $(=d-2)$ Hadamard code \cite{zhang2020proto,zhang2021}.
Supposing $r$ is even,
 $2^r-r-2$ Hadamard parity-bits are generated and attached to each H-CN as D1H-VNs.
The overall code rate of the PLDPC-Hadamard code is therefore
\begin{equation}
R = \frac{{n - m}}{{m \left( {{2^{r}} - {r}-2} \right)}  + n}.
\end{equation}
Throughout this paper, we assume that $d$ is even.
When $d$ is odd, $2^r-2$ Hadamard parity-bits are generated, and
the encoding and decoding algorithms become slightly different \cite{zhang2020proto,zhang2021}.

To speed up the convergence speed, a layered decoding algorithm has been proposed  \cite{Zhang2021layer}.
For $\alpha=0, 1,\ldots,M-1$ and $\beta=0, 1,\ldots,N-1$, we  denote
\begin{itemize}
\item $\mathcal{P}(\alpha)$ as the set of P-VNs connected to the $\alpha$-th H-CN;
\item $\mathcal{H}(\beta)$ as the set of H-CNs  connected to the $\beta$-th P-VN;
\item ${L}_{ch}^{\rm PVN}(\beta)$ as the channel log-likelihood-ratio (LLR) value of the $\beta$-th P-VN;
\item $\bm{L}_{ch}^{\rm D1H(\alpha)}$ as a vector consisting of the channel LLR values of the D1H-VNs connected to the $\alpha$-th H-CN;
\item ${L}_{app}^{\rm PVN}(\beta)$ as the \textit{a posteriori} probability (APP) LLR value of the $\beta$-th P-VN;
\item ${L}_{ex}^{\rm PVN}(\alpha,\beta)$ as the extrinsic LLR value from the $\beta$-th P-VN to the $\alpha$-th H-CN;
\item ${L}_{app}^{\rm H}(\alpha,\beta)$ as the APP LLR value computed by the $\alpha$-th H-CN for the $\beta$-th P-VN;
\item ${L}_{ex}^{\rm H}(\alpha,\beta)$ as the extrinsic LLR value sent from the $\alpha$-th H-CN   to the $\beta$-th P-VN.
\end{itemize}
After lifting the base matrix of a PLDPC-HC two times,
 the resultant adjacency matrix $\bm{H}_{M \times N}$ is divided into $m z_1$ layers (also called block rows),
 where each layer is composed of  $1 \times n z_1$ CPMs each of size $z_2 \times z_2$.
 Hence, each layer corresponds to a $z_2 \times n z_1 z_2$ matrix and contains $z_2$ H-CNs.
Since each H-CN connects $d$ P-VNs and $2^r-d$ D1H-VNs (when $r$ is even),
the $z_2$ H-CNs in one layer connects $dz_2$ P-VNs and $(2^r-d)z_2$ D1H-VNs.
 Table \ref{tb:one_layer} summarizes of the numbers of H-CNs, P-VNs and D1H-VNs contained in one layer.

\begin{table}[t]
\newcommand{\tabincell}[2]{\begin{tabular}{@{}#1@{}}#2\end{tabular}}
\centering\caption{Numbers of H-CNs, P-VNs and D1H-VNs contained in one layer when $r$ is even. $r = d - 2$. }\label{tb:one_layer} \footnotesize
\begin{center} 
 \begin{tabular}{|c|c|c|}
\hline
 No. of H-CNs & No. of P-VNs & No. of D1H-VNs   \\ 
 \hline
 $z_2$ & $dz_2$  & $(2^r-d)z_2$   \\
\hline
\end{tabular}
\end{center}
\end{table}
%
Defining $k$ as the layer number ($k=0,1,\ldots,m z_1 - 1$)
and $\mathcal{L}(k)=\{\alpha_{kz_2}, \alpha_{kz_2+1}, \ldots, \alpha_{kz_2+z_2 - 1}\}$ as the set of H-CNs in layer $k$,
the layered decoding algorithm
is described as follows \cite{Zhang2021layer}.
\begin{enumerate}
\item Initialization:
Set ${L}_{app}^{\rm PVN}(\beta)={L}_{ch}^{\rm PVN}(\beta)$  $\forall \beta$;
and set ${L}_{ex}^{\rm H}(\alpha,\beta) =0$ $\forall \alpha,\beta$.
\item \label{step:hld} Symbol maximum-a-posterior Hadamard sub-decoder: Set $k=0$.
\begin{enumerate}
\item \label{step:layer} For the $\alpha$-th H-CN in layer $k$ ($\alpha \in \mathcal{L}(k)$), perform the following computations.
\begin{enumerate}
\item  For $\beta  \in \mathcal{P}(\alpha)$, compute
        \begin{eqnarray}\label{eq:cpt_LexPVN2}
     {L}_{ex}^{\rm PVN}(\alpha,\beta) = {L}_{app}^{\rm PVN}(\beta) - {L}_{ex}^{\rm H}(\alpha,\beta)\cr \forall \beta \in \mathcal{P}(\alpha).
  \end{eqnarray}
  \item  \label{step:L_app^H} Compute ${L}_{app}^{\rm H}(\alpha,\beta)$ for the $\beta$-th P-VN ($\beta \in \mathcal{P}(\alpha)$)  using
\begin{eqnarray}\label{eq:cpt_LappH}
&&\!\!\!\!\!\!\!\!\!\!\!\!\!
 \bm{L}_{app}^{{\rm H}}(\alpha) = \{{L}_{app}^{\rm H}(\alpha,\beta): \beta \in \mathcal{P}(\alpha)\} \cr
&&\!\!\!\!\!\!\!\!\!\!\!\!\!
= \mathcal{T}\left[ \{ {L}_{ex}^{\rm PVN}(\alpha,\beta): \beta \in \mathcal{P}(\alpha) \},
         \bm{L}_{ch}^{\rm{D1H}(\alpha)} \right]
\end{eqnarray}
where $\mathcal{T}$ is a transformation involving the fast Hadamard transform (FHT) and the dual FHT (DFHT) operations \cite{Yue2007,zhang2020proto,zhang2021}.
\item  \label{step:update} Update  ${L}_{ex}^{\rm H}(\alpha,\beta)$ and ${L}_{app}^{\rm PVN}(\beta)$ using
\begin{eqnarray}
{L}_{ex}^{\rm H}(\alpha,\beta) &=& {L}_{app}^{\rm H}(\alpha,\beta) - {L}_{ex}^{\rm PVN}(\alpha,\beta);
 \label{eq:L_ex_H_layered} \cr
 && \makebox[2.2cm]{} \forall \beta \in \mathcal{P}(\alpha) \\ 
{L}_{app}^{\rm PVN}(\beta) &=& {L}_{app}^{\rm H}(\alpha,\beta);  \;\; \forall  \beta \in \mathcal{P}(\alpha).
 \label{eq:L_app_PVN_layered}
\end{eqnarray}
\end{enumerate}
\item If the last layer has not been reached, i.e., $k < m z_1 -1$, increment $k$ by $1$ and go to Step \ref{step:layer}).
\end{enumerate}
\item Repeat Step  \ref{step:hld}) $I$ times and make decisions on the P-VNs based on the sign of $L_{app}^{{\rm{PVN}}}\left( \beta \right)$  $\forall \beta$. 

\end{enumerate}
Note that the layered decoding algorithm neither returns any extrinsic information to
the D1H-VNs nor makes
hard decisions on the D1H-VNs. The algorithm only makes use of the channel information provided by the D1H-VNs
to aid the decoding of the PLDPC code and hence the P-VNs.

\section{Hardware Design of the Layered Decoder}\label{sect:hw_LDPCH}

This section presents and analyzes a hardware implementation of the layered decoding algorithm for PLDPC-HC.
First, we present the read and write operations of LLR values in random access memories (RAMs) \footnote{\color{black}  In a practical environment, the LLRs first are generated one-by-one by the demodulator at the receiver. They can then be passed to the next stage, i.e., decoder, one-by-one or in parallel (in a small number) and stored in the RAMs of the decoder. Thus the number of I/O interfaces between the demodulator and decoder can be designed to meet certain requirements.}
Second, we describe the pipeline structure of the symbol-maximum-a-posterior (symbol-MAP) Hadamard sub-decoder, which is composed mainly of FHT and DFHT components.
Third, we combine the RAMs and Hadamard sub-decoders
and propose a layered decoder architecture for PLDPC-HC.
Fourth, we analyze the decoding timing, latency and throughput of the proposed architecture.

\subsection{Read and Write Operations of RAMs}\label{sect:wr_rams}
As described in the layered decoding algorithm for the PLDPC-HC, there are
six types of LLRs. Among them $\{{L}_{ex}^{\rm PVN}(\alpha,\beta) \}$ in \eqref{eq:cpt_LexPVN2}
and $\{\bm{L}_{app}^{\rm H}(\alpha,\beta) \}$   in \eqref{eq:cpt_LappH}
are only temporary values in the computation process and need not to be stored, whereas
the other four types of LLRs, i.e., $\{{L}_{ch}^{\rm PVN}(\beta)\}$, $\{{L}_{app}^{\rm PVN}(\beta)\}$, $\{{L}_{ex}^{\rm H}(\alpha,\beta)\}$ and $\{{\bm L}_{ch}^{\rm D1H(\alpha)}\}$, are not temporary and thus
need to be stored in RAMs. 

Referring to Table \ref{tb:one_layer}, the $z_2$ H-CNs in each layer
connect $d z_2$ P-VNs and $(2^r-d) z_2$ D1H-VNs (when $r$ is even).
Using the layered decoding algorithm to process each layer,
we therefore need to retrieve $d z_2$ values of $\{{L}_{ch}^{\rm PVN}(\beta)\}$
(during initialization)
 or $d z_2$ values of $\{{L}_{app}^{\rm PVN}(\beta)\}$
 (to be used in \eqref{eq:cpt_LexPVN2}); and $z_2$ vectors of $\{{\bm L}_{ch}^{\rm D1H(\alpha)}\}$.
Note that each vector of $\{{\bm L}_{ch}^{\rm D1H(\alpha)}\}$ contains $2^r-d$ LLR values.
According to \eqref{eq:cpt_LexPVN2}, we also need to retrieve $d z_2$ values of $\{{L}_{ex}^{\rm H}(\alpha,\beta)\}$
in order to compute the $d z_2$ values of $\{{L}_{ex}^{\rm PVN}(\alpha,\beta)\}$.
In our design, we form sets of LLRs where each set has a size of $z_2$ --- the same size as the second lifting factor.
For a PLDPC-HC with an $m \times n$ protomatrix and lifting factors $z_1$ and $z_2$,
$\{{L}_{ch}^{\rm PVN}(\beta)\}$ will be divided into $N/z_2 = n z_1$ sets,
$\{{L}_{app}^{\rm PVN}(\beta)\}$ into $N/z_2 = n z_1$ sets,
$\{{L}_{ex}^{\rm H}(\alpha,\beta)\}$ into $Md/z_2 = m d z_1$ sets,
and $\{{\bm L}_{ch}^{\rm D1H(\alpha)}\}$ into $M/z_2=mz_1$ sets.

\begin{figure}[t]
\centerline{
\includegraphics[width=.7\columnwidth]{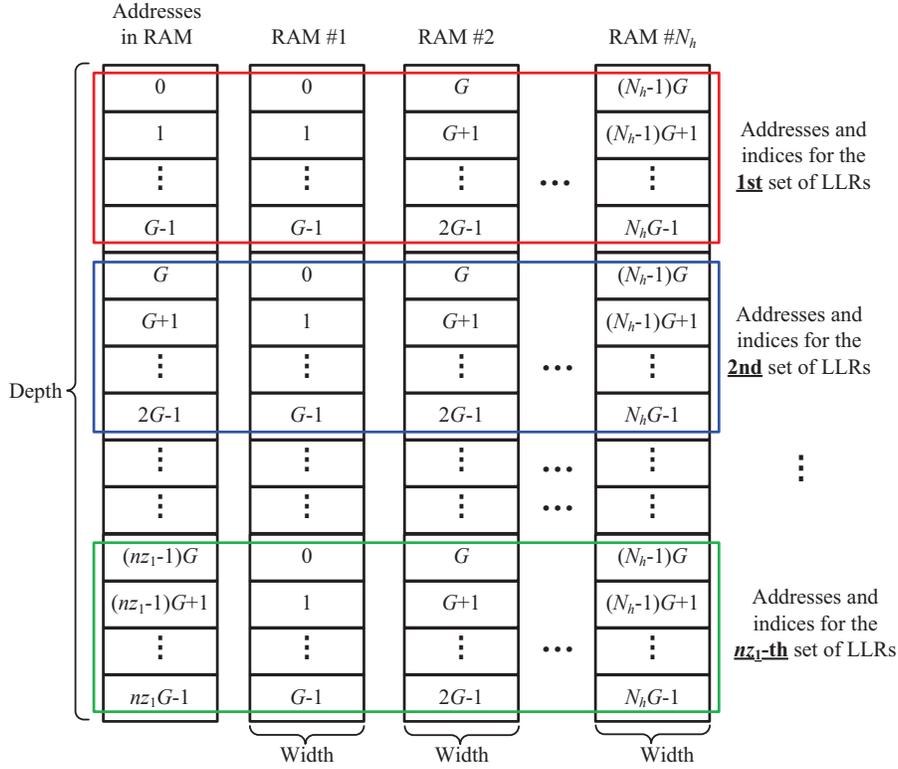}}
 \caption{ RAM arrangement for $\{{L}_{ch}^{\rm PVN}(\beta)\}$ or $\{{L}_{app}^{\rm PVN}(\beta)\}$ corresponding P-VNs. 
 }
  \label{fig:RAM}
\end{figure}

To achieve reading/retrieving $N_h$ data from memories in one clock cycle, we use $N_h$ RAMs to store each type of LLRs, where $0 < N_h \le z_2$ and $G = z_2/N_h$ is an integer and represents the number of groups.
Taking $\{{L}_{ch}^{\rm PVN}(\beta)\}$ which is related to the
P-VNs as an example, each set of LLRs, i.e., a total of $z_2$ LLR values, is further divided into $G$ groups.
Referring to Fig. \ref{fig:RAM}, the addresses $0, 1, \ldots, G-1$ in
the $N_h$ RAMs are to store the first set of ${L}_{ch}^{\rm PVN}(\beta)$.
In particular, RAM \#1 stores the first group of LLRs, i.e., LLRs with indices $0, 1, \ldots, G-1$;
 RAM \#2 stores the second group of LLRs, i.e., LLRs with indices $G, G+1, \ldots, 2G-1$;
 $\ldots$; and RAM \#$N_h$ stores the $N_h$-th group of LLRs, i.e.,
 LLRs with indices $(N_h-1)G,(N_h-1)G+1,\ldots, N_h G-1$.
Using a similar fashion, the addresses $G, G+1, \ldots, 2G-1$ in the
$N_h$ RAMs are to store the second set of ${L}_{ch}^{\rm PVN}(\beta)$.
The arrangement is repeated
 until all $n z_1$ sets of ${L}_{ch}^{\rm PVN}(\beta)$ are stored in the $N_h$ RAMs.

With the above storage arrangement, in each clock cycle
$N_h$ values from the same LLR set can be retrieved from the $N_h$ RAMs.
Using Fig. \ref{fig:RAM} as an example,
at clock $t = 1$ the $N_h$ LLR values stored at Address \#0 (with indices $0, G, \ldots, (N_h-1)G$) are retrieved;
at clock $t = 2$, the $N_h$ LLR values stored at Address \#1 (with indices $1, G+1, \ldots, (N_h-1)G+1$)  are retrieved;
$\cdots$;
at clock $t = G$, the $N_h$ LLR values stored at Address \#$G-1$  (with indices  $G-1, 2G-1, \ldots, N_hG-1$)  are retrieved.
Thus one set of LLR values (i.e., $z_2$ LLR values) can be retrieved in $G$ clock cycles.
Hence, reading or writing $d$ sets of ${L}_{ch}^{\rm PVN}(\beta)$ or ${L}_{app}^{\rm PVN}(\beta)$ for each layer
requires $dG$ clock cycles.

\begin{figure}[t]
\centerline{
\includegraphics[width=0.7\columnwidth]{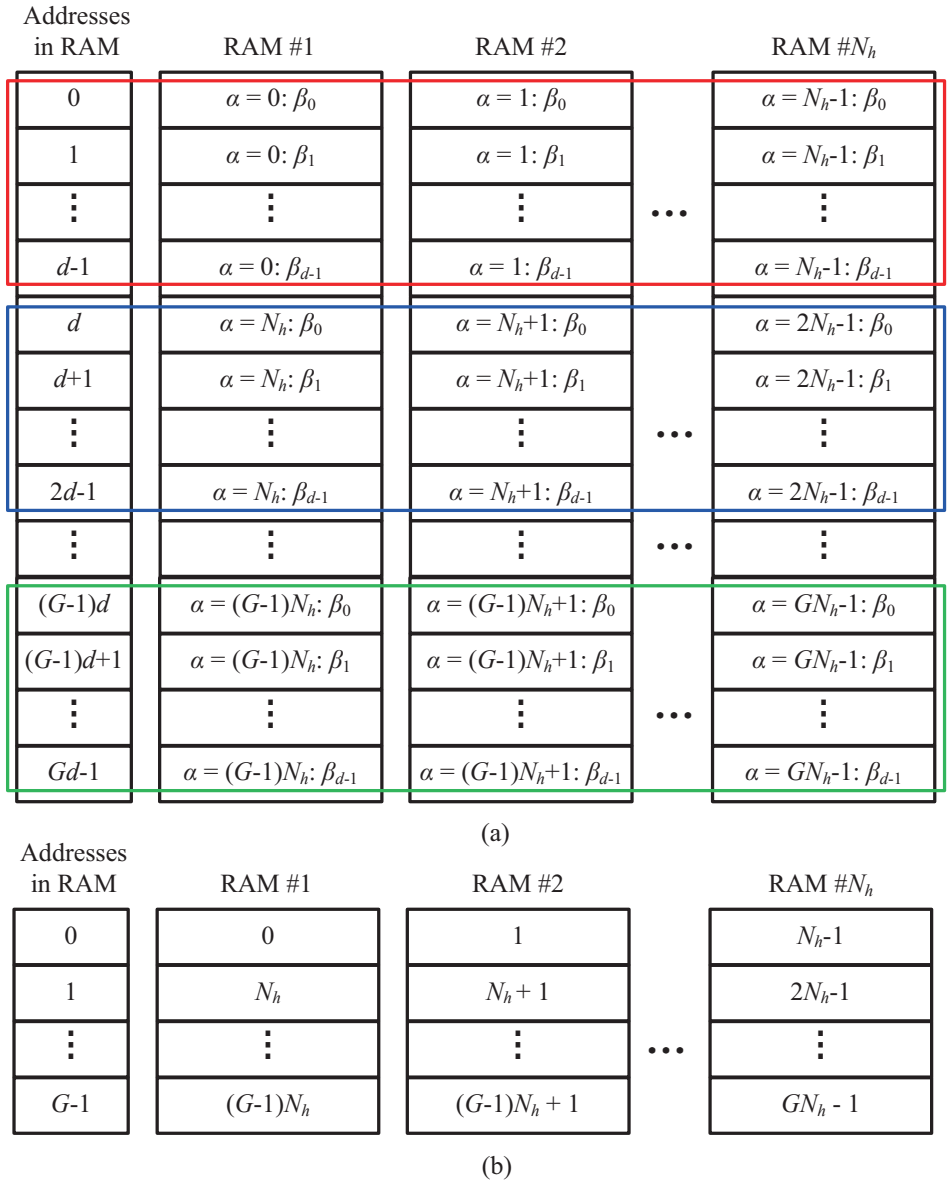}}
 \caption{Storage arrangement of (a) $\{{L}_{ex}^{\rm H}(\alpha,\beta)\}$ and (b) $\{{\bm L}_{ch}^{\rm D1H(\alpha)}\}$  in $N_h$ RAMs. The first layer of H-CNs is being considered.}
  \label{fig:H-RAM}
\end{figure}

We use a similar storage arrangement for $\{{L}_{ex}^{\rm H}(\alpha,\beta)\}$  and $\{{\bm L}_{ch}^{\rm D1H(\alpha)}\}$, which correspond to H-CNs and D1H-VNs, respectively. The only differences are that the RAMs will have different depths and widths.
Fig. \ref{fig:H-RAM} shows the storage arrangement of
 $\{{L}_{ex}^{\rm H}(\alpha,\beta)\}$ and $\{{\bm L}_{ch}^{\rm D1H(\alpha)}\}$ (corresponding to the first layer) in $N_h$ RAMs.
In Fig. \ref{fig:H-RAM}(a),
``$\alpha = i: \beta_j$'' ($i=0,\ldots,GN_h -1; j=0,\ldots,d-1$) denotes the
$j$-th H-CN connected to the $i$-th P-VN; and hence
``$\alpha = i: \beta_0$'' to ``$\alpha = i: \beta_{d-1}$''
represent $\mathcal{P}(\alpha = i)$, i.e., all the
P-VNs connected to the $i$-th H-CN.
In Fig. \ref{fig:H-RAM}(b), each address stores
the $2^r-d$ channel LLRs corresponding to
the $2^r-d$ D1H-VNs connected to the same H-CN.

{\color{black} \textit{Remark:}  The aforementioned arrangement of the LLRs in the RAMs is valid regardless of single-port RAMs or dual-port RAMs being used.  
In other words,
no read/write conflicts will occur whether single-port RAMs or dual-port RAMs are used.
In the actual hardware implementation, we use
dual-port RAMs instead of single-port ones.
Since two memory locations in each dual-port RAM can be accessed (read and/or write) at the same time,
the number of clock cycles required to read/write one set of LLRs
can be further reduced by half compared with the discussion above.
The theoretical latency and throughput derived in Sect.~\ref{subsect:l&t} and the experimental results shown in Sect.~\ref{sect:results} are all based on the use dual-port RAMs.}

Supposing we have retrieved $N_h$ values for $\{{L}_{ch}^{\rm PVN}(\beta)\}$ or $\{{L}_{app}^{\rm PVN}(\beta)\}$,
we need to interleave them --- a process similar to that used in QC-LDPC decoding \cite{Sham2013}.
For each layer, the exact connections between the H-CNs and the P-VNs are determined by the CPMs,
and hence the interleaver can be realized by a simple cyclic shifter.
Assuming that the offset value of a CPM equals $p$ $(0 \le p < z_2)$,
 we calculate the quotient $q_u = \lfloor p/G \rfloor$ and the remainder $r_e = p \mod G$, where $\lfloor x \rfloor$ denotes the greatest integer less than or equal to $x$ and ``mod'' denotes the modulus operation.
When $(address \mod G) < r_e$, the corresponding $N_h$ LLRs are cyclically shifted to the left by $(q_u+1 \mod N_h)$;
otherwise, these LLRs are cyclically shifted to the left by $q_u$.

\begin{figure}[t]
\centerline{
\includegraphics[width=1\columnwidth]{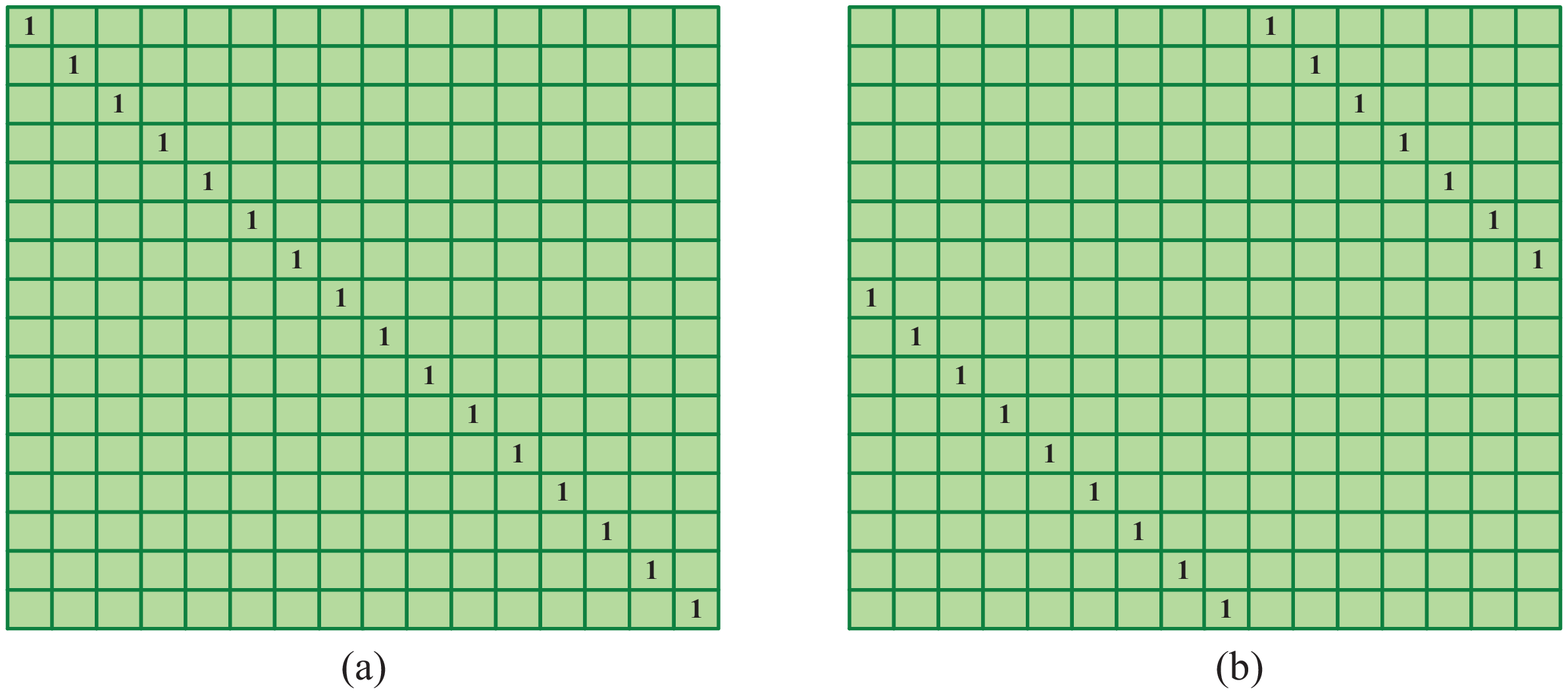}}
 \caption{ (a) A $16 \times 16$ identity matrix, i.e., a $16 \times 16$ circulant permutation matrix (CPM) with $p = 0$; (b) A $16 \times 16$ CPM with $p = 9$, which can be obtained by cyclically shifting the $16 \times 16$ identity matrix to the right by $p = 9$ columns. }
  \label{fig:cpm}
\end{figure}

{\underline{Example}}:
We assume that $z_2 = 16$.
Fig. \ref{fig:cpm}(a) shows a $16 \times 16$ identity matrix, i.e., a $16 \times 16$ CPM with $p = 0$;
Fig. \ref{fig:cpm}(b) depicts a $16 \times 16$ CPM with $p=9$, which can be obtained by cyclically shifting the $16 \times 16$ identity matrix to the right by $9$ columns.
Assume that the CPM with $p=9$ corresponds to one set of ${L}_{ch}^{\rm PVN}(\beta)$ with
 indices $[0\ 1\ 2\ 3\ 4\ 5\ 6\ 7\ 8\ 9\ 10\ 11\ 12\ 13\ 14\ 15]$.
After retrieving these LLRs from the RAMs and interleaving them,
these indices are expected to be re-ordered into $[9\ 10\ 11\ 12\ 13\ 14\ 15\ 0\ 1\ 2\ 3\ 4\ 5\ 6\ 7\ 8]$.
Suppose we use $N_h=4$ RAMs to store this set of ${L}_{ch}^{\rm PVN}(\beta)$.
According to our aforementioned storage scheme,
each set of ${L}_{ch}^{\rm PVN}(\beta)$ is divided into $G=z_2/N_h = 4$ groups;
and each RAM would use the first $G=4$ addresses, i.e., Addresses $\#0, \#1, \#2, \#3$, to store $4$ LLR values.
The storage arrangement is shown in Table \ref{tb:example_ram}.
As $p=9$, we have $q_u = \lfloor p/G \rfloor = 2$ and $r_e = p \mod G = 1$.
Once the $N_h=4$ LLRs are retrieved, we process them as follows.
\begin{itemize}
\item Cyclically shift the LLRs stored at Address \#0 ($< r_e = 1$), i.e., LLRs with indices $[0\ 4\ 8\ 12]$, to the left by $q_u+1 \mod N_h = 3$ and
the order of the indices becomes $[12\ 0\ 4\ 8]$;
\item Cyclically shift the LLRs  stored at Address \#1 ($\ge r_e = 1$) to the left by $q_u = 2$ and
the order of the indices becomes $[9\ 13\ 1\ 5]$;
\item Cyclically shift the LLRs  stored at Address \#2 ($\ge r_e = 1$) to the left by $q_u = 2$ and
the order of the indices becomes $[10\ 14\ 2\ 6]$;
\item Cyclically shift the LLRs  stored at Address \#3 ($\ge r_e = 1$) to the left by $q_u = 2$ and
the order of the indices becomes $[11\ 15\ 3\ 7]$.
\end{itemize}
Therefore, the expected interleaving effect $[9\ 10\ 11\ 12\ 13\ 14\ 15\ 0\ 1\ 2\ 3\ 4\ 5\ 6\ 7\ 8]$ can be achieved by such a process.
Note that the ``write'' operation can be regarded as the reverse process of the ``read'' operation.
Hence the procedures are similar and are omitted here.

\begin{table}[t]
\newcommand{\tabincell}[2]{\begin{tabular}{@{}#1@{}}#2\end{tabular}}
\centering\caption{Example for storing a set of LLRs in $N_h=4$ RAMs. $z_2 = 16$ and $G = z_2 / N_h = 4$. }\label{tb:example_ram} \footnotesize
\begin{center} 
 \begin{tabular}{|c||c|c|c|c|}
\hline
\multirow{2}{*}{Address} &\multicolumn{4}{c|}{Indices} \\ \cline{2-5}
& RAM \#$1$  & RAM \#$2$  & RAM \#$3$   & RAM \#$4$   \\
\hline
$0$& $0$  & $4$  & $8 $ & $12$  \\
\hline
$1$& $1$  & $5$  & $9 $ & $13$  \\
\hline
$2$& $2$  & $6$  & $10 $ & $14$  \\
\hline
$3$& $3$  & $7$  & $11 $ & $15$  \\
\hline
\end{tabular}
\end{center}
\end{table}

\subsection{Operation of A Symbol-MAP Hadamard Sub-decoder}\label{subsec:str_h}
The Hadamard sub-decoder can be considered as the kernel of the PLDPC-HC layered decoder in our implementation and hence will be described in detail.
For an order-$r$ Hadamard code, the corresponding Hadamard
matrices of size $q \times q$ can be recursively constructed by
\begin{eqnarray}\label{eq:H}
{\pm\bm{H}_q} &=& \{\pm \bm{h}_j, j = 0,1,\ldots, q-1\} \nonumber\\
 &=&  \left[ {\begin{array}{*{20}{c}}
{ \pm {\bm{H}_{q/2}}}&{ \pm {\bm{H}_{q/2}}}\\
{ \pm {\bm{H}_{q/2}}}&{ \mp {\bm{H}_{q/2}}}
\end{array}} \right]
\end{eqnarray}
where $q = 2^r$ equals the code length and $\pm\bm{H}_1 = [\pm1]$.
Each column $\pm \bm{h}_j$ of the Hadamard matrices corresponds to a Hadamard codeword, and hence there
is a total of $2q = 2^{r+1}$ codewords in $\pm\bm{H}_q$.
In \eqref{eq:H16}, we show
the $16 \times 16$ Hadamard matrices
$\pm \bm{H}_{16}$ corresponding to the order-$r=4$ Hadamard code having $2^{r+1} = 32$ codewords.
Note that the codewords are formed by mapping each $+1$ in the Hadamard matrices to bit ``$0$'' and
each $-1$ to bit ``$1$''.

\begin{figure*}[t]
\begin{equation}\label{eq:H16}
\pm \bm{H}_{16} = \left[ {\begin{array}{*{20}{c}}
{\color{black}{ \pm 1}}&{\color{black}{ \pm 1}}&{\color{black}{ \pm 1}}&{\color{black}{ \pm 1}}&{\color{black}{ \pm 1}}&{\color{black}{ \pm 1}}&{\color{black}{ \pm 1}}&{\color{black}{ \pm 1}}&{\color{black}{ \pm 1}}&{\color{black}{ \pm 1}}&{\color{black}{ \pm 1}}&{\color{black}{ \pm 1}}&{\color{black}{ \pm 1}}&{\color{black}{ \pm 1}}&{\color{black}{ \pm 1}}&{\color{black}{ \pm 1}}\\
{\color{black} \pm 1}&{\color{black} \mp 1}&{\color{black} \pm 1}&{\color{black} \mp 1}&{\color{black} \pm 1}&{\color{black} \mp 1}&{\color{black} \pm 1}&{\color{black} \mp 1}&{\color{black} \pm 1}&{\color{black} \mp 1}&{\color{black} \pm 1}&{\color{black} \mp 1}&{\color{black} \pm 1}&{\color{black} \mp 1}&{\color{black} \pm 1}&{\color{black} \mp 1}\\
{\color{black} \pm 1}&{\color{black} \pm 1}&{\color{black} \mp 1}&{\color{black} \mp 1}&{\color{black} \pm 1}&{\color{black} \pm 1}&{\color{black} \mp 1}&{\color{black} \mp 1}&{\color{black} \pm 1}&{\color{black} \pm 1}&{\color{black} \mp 1}&{\color{black} \mp 1}&{\color{black} \pm 1}&{\color{black} \pm 1}&{\color{black} \mp 1}&{\color{black} \mp 1}\\
{ \pm 1}&{ \mp 1}&{ \mp 1}&{ \pm 1}&{ \pm 1}&{ \mp 1}&{ \mp 1}&{ \pm 1}&{ \pm 1}&{ \mp 1}&{ \mp 1}&{ \pm 1}&{ \pm 1}&{ \mp 1}&{ \mp 1}&{ \pm 1}\\
{\color{black} \pm 1}&{\color{black} \pm 1}&{\color{black} \pm 1}&{\color{black} \pm 1}&{\color{black} \mp 1}&{\color{black} \mp 1}&{\color{black} \mp 1}&{\color{black} \mp 1}&{\color{black} \pm 1}&{\color{black} \pm 1}&{\color{black} \pm 1}&{\color{black} \pm 1}&{\color{black} \mp 1}&{\color{black} \mp 1}&{\color{black} \mp 1}&{\color{black} \mp 1}\\
{ \pm 1}&{ \mp 1}&{ \pm 1}&{ \mp 1}&{ \mp 1}&{ \pm 1}&{ \mp 1}&{ \pm 1}&{ \pm 1}&{ \mp 1}&{ \pm 1}&{ \mp 1}&{ \mp 1}&{ \pm 1}&{ \mp 1}&{ \pm 1}\\
{ \pm 1}&{ \pm 1}&{ \mp 1}&{ \mp 1}&{ \mp 1}&{ \mp 1}&{ \pm 1}&{ \pm 1}&{ \pm 1}&{ \pm 1}&{ \mp 1}&{ \mp 1}&{ \mp 1}&{ \mp 1}&{ \pm 1}&{ \pm 1}\\
{ \pm 1}&{ \mp 1}&{ \mp 1}&{ \pm 1}&{ \mp 1}&{ \pm 1}&{ \pm 1}&{ \mp 1}&{ \pm 1}&{ \mp 1}&{ \mp 1}&{ \pm 1}&{ \mp 1}&{ \pm 1}&{ \pm 1}&{ \mp 1}\\
{\color{black} \pm 1}&{\color{black} \pm 1}&{\color{black} \pm 1}&{\color{black} \pm 1}&{\color{black} \pm 1}&{\color{black} \pm 1}&{\color{black} \pm 1}&{\color{black} \pm 1}&{\color{black} \mp 1}&{\color{black} \mp 1}&{\color{black} \mp 1}&{\color{black} \mp 1}&{\color{black} \mp 1}&{\color{black} \mp 1}&{\color{black} \mp 1}&{\color{black} \mp 1}\\
{ \pm 1}&{ \mp 1}&{ \pm 1}&{ \mp 1}&{ \pm 1}&{ \mp 1}&{ \pm 1}&{ \mp 1}&{ \mp 1}&{ \pm 1}&{ \mp 1}&{ \pm 1}&{ \mp 1}&{ \pm 1}&{ \mp 1}&{ \pm 1}\\
{ \pm 1}&{ \pm 1}&{ \mp 1}&{ \mp 1}&{ \pm 1}&{ \pm 1}&{ \mp 1}&{ \mp 1}&{ \mp 1}&{ \mp 1}&{ \pm 1}&{ \pm 1}&{ \mp 1}&{ \mp 1}&{ \pm 1}&{ \pm 1}\\
{ \pm 1}&{ \mp 1}&{ \mp 1}&{ \pm 1}&{ \pm 1}&{ \mp 1}&{ \mp 1}&{ \pm 1}&{ \mp 1}&{ \pm 1}&{ \pm 1}&{ \mp 1}&{ \mp 1}&{ \pm 1}&{ \pm 1}&{ \mp 1}\\
{ \pm 1}&{ \pm 1}&{ \pm 1}&{ \pm 1}&{ \mp 1}&{ \mp 1}&{ \mp 1}&{ \mp 1}&{ \mp 1}&{ \mp 1}&{ \mp 1}&{ \mp 1}&{ \pm 1}&{ \pm 1}&{ \pm 1}&{ \pm 1}\\
{ \pm 1}&{ \mp 1}&{ \pm 1}&{ \mp 1}&{ \mp 1}&{ \pm 1}&{ \mp 1}&{ \pm 1}&{ \mp 1}&{ \pm 1}&{ \mp 1}&{ \pm 1}&{ \pm 1}&{ \mp 1}&{ \pm 1}&{ \mp 1}\\
{ \pm 1}&{ \pm 1}&{ \mp 1}&{ \mp 1}&{ \mp 1}&{ \mp 1}&{ \pm 1}&{ \pm 1}&{ \mp 1}&{ \mp 1}&{ \pm 1}&{ \pm 1}&{ \pm 1}&{ \pm 1}&{ \mp 1}&{ \mp 1}\\
{\color{black} \pm 1}&{\color{black} \mp 1}&{\color{black} \mp 1}&{\color{black} \pm 1}&{\color{black} \mp 1}&{\color{black} \pm 1}&{\color{black} \pm 1}&{\color{black} \mp 1}&{\color{black} \mp 1}&{\color{black} \pm 1}&{\color{black} \pm 1}&{\color{black} \mp 1}&{\color{black} \pm 1}&{\color{black} \mp 1}&{\color{black} \mp 1}&{\color{black} \pm 1}
\end{array}} \right]
\end{equation}
\end{figure*}

When the Hadamard order $r$ is even, it has been proven that
there always exists a length-$d= r+2$ single-parity-check (SPC) codeword
``embedded'' in each Hadamard codeword \cite{Yue2007,zhang2020proto,zhang2021}, i.e.,
\begin{eqnarray}
\label{eq:SPC}\label{eq:spc}
[ \pm h_{0,j}  \oplus \pm h_{1,j} \oplus \cdots\oplus  {\ \pm h_{2^{k-1},j}} \oplus \cdots  \oplus &\cr
&\!\!\!\!\!\!\!\!\!\!\!\!\!\!\!\!\!\!\!\!\!\!\!\!\!\!\!\!\!\!\!\!\!\!\!\!\!\!\!\!\!\!\!\!\! {\ \pm h_{2^{r-1},j}} ]\ \oplus\ \pm h_{2^r-1,j}   = 0,
\end{eqnarray}
where the symbol $\oplus$  represents the XOR operator.
(In \eqref{eq:H16}, the length-$6$ SPC constraint is $\pm h_{0,j}\oplus \pm h_{1,j}\oplus  \pm h_{2,j}\oplus   \pm h_{4,j}\oplus  \pm h_{8,j}\oplus  \pm h_{15,j} = 0 \; \forall \; j$ and the corresponding $6$ bits are marked in red color.)
In each H-CN of the PLDPC-HC described in Section \ref{sect:str_LDPCH}, the length-$d$ SPC codeword
is formed by the $d$ P-VNs to which the H-CN is connected.
Using these $d$ bits as inputs to the Hadamard encoder,
$2^r-d$ Hadamard parity-check bits corresponding to the D1H-VNs attached to the H-CN can be generated.
 (In the case of an order-$4$ Hadamard code, $d=6$ bits are input to the Hadamard encoder which generates $2^r- d = 10$ Hadamard parity-check bits.)

To decode Hadamard codes, a symbol-MAP decoding algorithm has been proposed
 \cite{Yue2007,zhang2020proto,zhang2021}.
We define
\begin{eqnarray}
\bm{L}_{ch}^{\rm H} &=& [{L}_{ch}^{\rm H}(0) \; {L}_{ch}^{\rm H}(1)  \; \cdots  \; {L}_{ch}^{\rm H}(2^r-1)] ^T, \\
\bm{L}_{apr}^{\rm H} &=& [{L}_{apr}^{\rm H}(0) \; {L}_{apr}^{\rm H}(1) \;  \cdots \;  {L}_{apr}^{\rm H}(2^r-1)] ^T,\\
\bm{L}_{app}^{\rm H} &=& [{L}_{app}^{\rm H}(0) \; {L}_{app}^{\rm H}(1) \;  \cdots \;  {L}_{app}^{\rm H}(2^r-1)] ^T,
\end{eqnarray}
as the channel, the \textit{a priori} and the \textit{a posteriori} LLR information of the coded bit, respectively.
Note that $\bm{L}_{ch}^{\rm H}$ contains only $2^r-d$ channel observations coming from the D1H-VNs while the remaining $d$ values are set to $0$. On the other hand,
$\bm{L}_{apr}^{\rm H}$ has $d$ non-zero values coming from P-VNs (i.e., repeat decoder) while the remaining $2^r-d$ values are set to $0$.
(Please refer to \cite[Section III-B]{zhang2020proto} and \cite[Section III-B]{zhang2021}
for the detailed arrangement of $\bm{L}_{ch}^{\rm H}$ and $\bm{L}_{apr}^{\rm H}$.)
Based on $\bm{L}_{ch}^{\rm H}$ and $\bm{L}_{apr}^{\rm H}$, $L_{app}^{\rm H}(i)$
is computed using
\begin{eqnarray}\label{eq:cpt_app}
L_{app}^{\rm H}(i)
&=& \ln \frac{{\sum\limits_{\pm H\left[ {i,j} \right] =  + 1} {\gamma \left( { \pm {\bm{h}_j}} \right)} }}
{{\sum\limits_{\pm H\left[ {i,j} \right] =  - 1} {\gamma \left( { \pm {\bm{h}_j}} \right)} }},
\end{eqnarray}
\noindent
where $\gamma \left( \pm {\bm{h}_j} \right) = \exp \left( {\left\langle { \pm {\bm{h}_j}, \bm{L}_{ch}^{\rm H}+\bm{L}_{apr}^{\rm H}} \right\rangle } / 2 \right)$
represents the \textit{a posteriori} ``information'' of the codeword $\pm\bm{h}_j$;
and $\left\langle  \cdot  \right\rangle $ denotes the inner-product operator.
\begin{figure}[t]
\centerline{
\includegraphics[width=0.7\columnwidth]{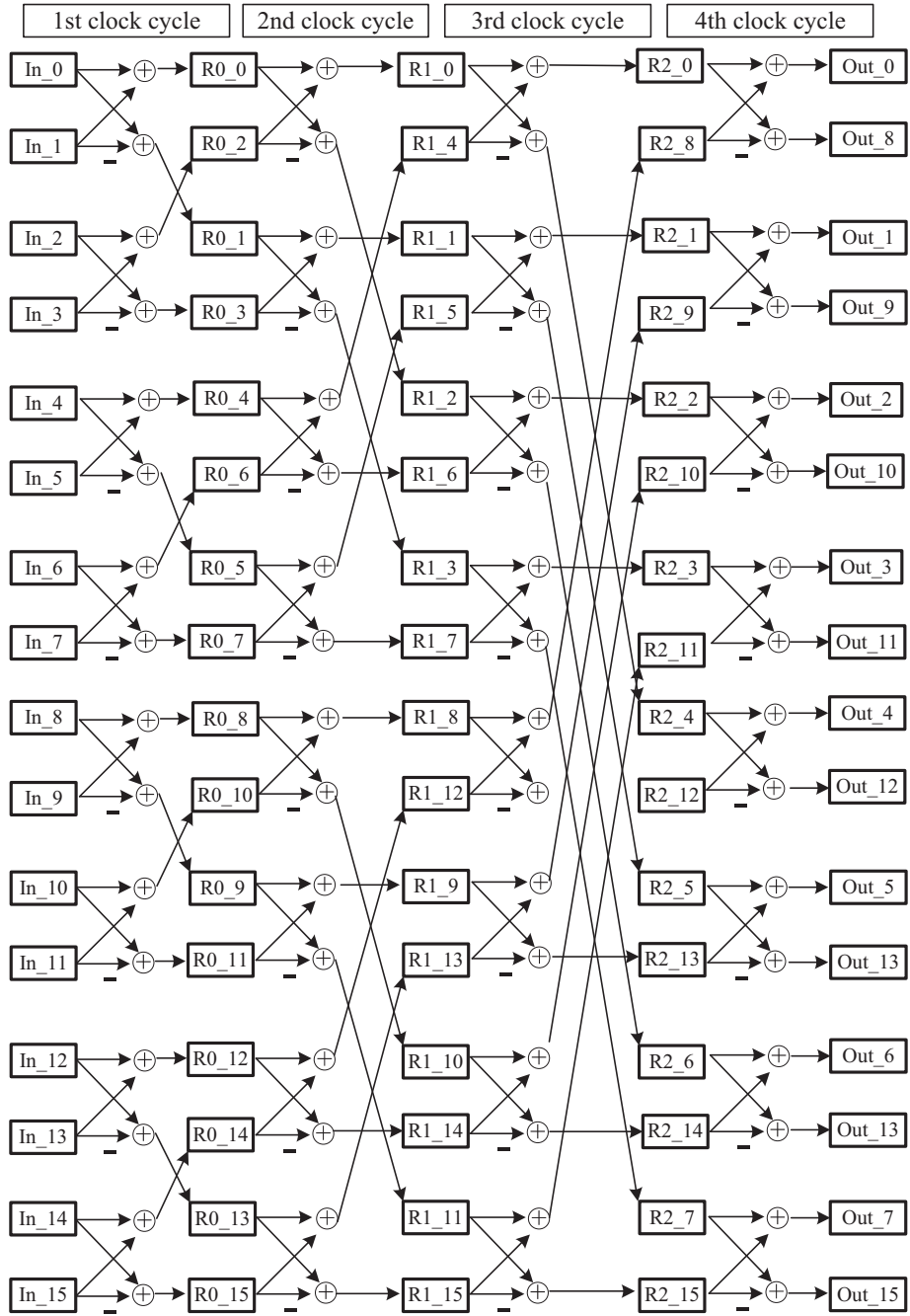}}
 \caption{Pipeline structure of a FHT block for $r = 4$ \cite{Yue2007}. {\color{black}Connections to the clock  are omitted for clarity.} }
  \label{fig:FHT}
\end{figure}
\begin{figure*}[t]
\centerline{
\includegraphics[width=1\columnwidth]{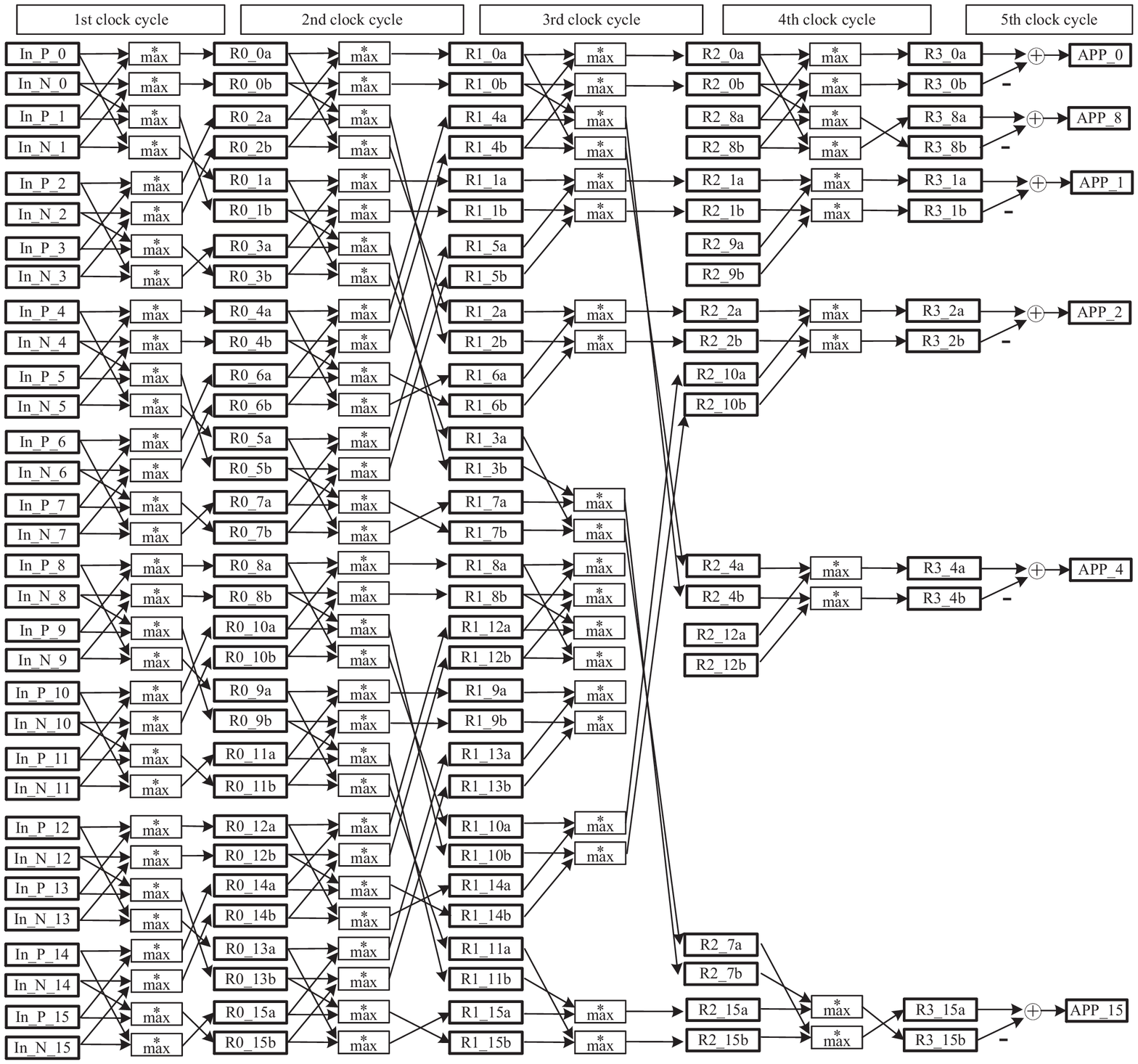}}
 \caption{Pipeline structure of a reduced DFHT block for $r = 4$. An additional clock cycle (5-th clock cycle) is shown
 for the computation of APP LLRs ${L}_{app}^{\rm H}(i)$ \cite{Yue2007}.
Connections to the clock  are omitted for clarity.}
  \label{fig:DFHT}
\end{figure*}
Since the Hadamard matrix has a butterfly-like structure, our Hadamard decoder design is based on the fast Hadamard transform (FHT) block and the dual FHT (DFHT) block \cite{Li2003,Yue2007,thc:2020}.
\begin{enumerate}
\item
We first use a FHT block to compute ${\left\langle { + {\bm{h}_j}, \bm{L}_{ch}^{\rm H}+\bm{L}_{apr}^{\rm H}} \right\rangle }$. Using the structure of the FHT block for $r=4$ shown in Fig. \ref{fig:FHT} as an example, the
inputs are ${\rm{In}}\_j = L^{\rm H}_{ch}(j)+L^{\rm H}_{apr}(j)$ and the outputs are ${\rm{Out}}\_j = 2  \ln\left[\gamma \left( + {\bm{h}_j} \right)\right]$ ($j=0,1,\ldots, 15$).
Then,
$ \ln\left[\gamma \left( + {\bm{h}_j} \right)\right]$
is readily obtained from $2  \ln\left[\gamma \left( + {\bm{h}_j} \right)\right]$
by shifting the least significant bit out. Moreover,
$ \ln\left[\gamma \left( - {\bm{h}_j} \right)\right]$ is readily available because
$ \ln\left[\gamma \left( - {\bm{h}_j} \right)\right] = - \ln\left[\gamma \left( + {\bm{h}_j} \right)\right]$.
There are $r=4$ stages in the FHT block and thus a latency of $r=4$ clock cycles is required.

\item The structure of a DFHT block is similar to that of a FHT block, but with twice the number of inputs and outputs.
Using the structure of the DFHT block for $r=4$ shown in Fig. \ref{fig:DFHT} as an example,
the inputs to the DFHT block are
$ \ln\left[\gamma \left( + {\bm{h}_j} \right)\right]$ and $ \ln\left[\gamma \left( - {\bm{h}_j} \right)\right]$; and the outputs are $\ln \left[ \sum_{\pm H\left[ {i,j} \right] =  + 1} {\gamma \left( { \pm {\bm{h}_j}} \right)} \right]$ and $\ln \left[ \sum_{\pm H\left[ {i,j} \right] =  - 1} {\gamma \left( { \pm {\bm{h}_j}} \right)} \right]$ ($j=0,1,\ldots, 15$).
{\color{black} The module $ \overset{*}{\rm max}$ in the DFHT block represents 
the Jacobian logarithm, i.e.,}
 \begin{eqnarray}
\overset{*}{\max}(a,b) &=& \ln [\exp(a) + \exp(b)] \nonumber\\
&=& \max(a,b) + \ln\left[1 + \exp( - \left| {a - b} \right|)\right]\nonumber\\
\end{eqnarray}
where $\max(a,b)$ returns the greater value between $a$ and $b$.
In our design, we use a comparison operation to realize $\max(a,b)$, a look-up-table to realize $\ln\left[1 + \exp( - \left| {a - b} \right|)\right]$ and an addition operation to sum the above outputs.


As we only need to feedback values related to the $r+2$ information bits, the structure of DFHT block can be further simplified to minimize resources requirement.
Same as the FHT block, the DFHT block contains $r$ stages and thus has a latency of $r$ clock cycles.
\end{enumerate}
Finally, for $i=0,1,\ldots, 2^{k-1}, \ldots, 2^{r-1}, 2^r-1$, it takes another clock cycle to compute
\begin{itemize}
\item the $r+2$ LLR values ${L}_{app}^{\rm H}(i)$ which equals
$$\ln \left[ \sum_{\pm H\left[ {i,j} \right] =  + 1} {\gamma \left( { \pm {\bm{h}_j}} \right)} \right] - \ln \left[ \sum_{\pm H\left[ {i,j} \right] = - 1} {\gamma \left( { \pm {\bm{h}_j}} \right)} \right],$$
\item the $r+2$ extrinsic LLR messages $L_{ex}^{\rm H}(i)$ which is computed using \eqref{eq:L_ex_H_layered}. 
\end{itemize}

Overall, it takes $2r+1$ clock cycles to complete one set of computation.
Note that the FHT and DFHT blocks have pipeline structures, and the results computed in each stage will be stored in registers.
{\color{black}To simplify the presentations of the structures of the
FHT block (Fig. \ref{fig:FHT}) and the DFHT block (Fig. \ref{fig:DFHT}), we omit all connections to the clock in the figures.
}

\subsection{Layered Decoder Architecture}\label{subsec:arch}
\begin{figure}[t]
\centerline{
\includegraphics[width=0.6\columnwidth]{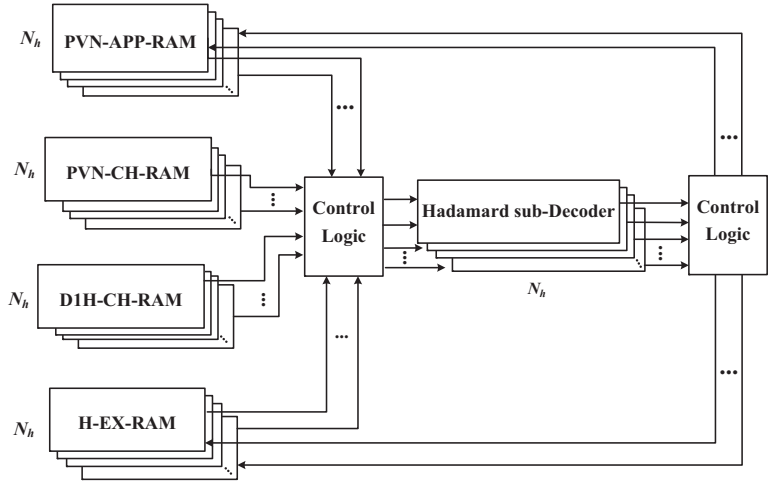}}
 \caption{Proposed layered PLDPC-Hadamard decoder with four types of RAMs, Hadamard
 sub-decoders and control logics.}
  \label{fig:PLDPCH_DEC}
\end{figure}

Referring to Fig.~\ref{fig:PLDPCH_DEC}, we propose an architecture of PLDPC-Hadamard layered decoder
based on RAMs and Hadamard sub-decoders.
Moreover, we assume that there are $N_h$ Hadamard sub-decoders.
In addition to RAMs and sub-decoders, the architecture contains control logics.
The control logics are dependent on the structure of the
adjacency matrix which has a relatively simple quasi-cyclic format.
They are used to ensure that the correct data are loaded into the individual Hadamard sub-decoder and the updated data are written to the correct memory locations
\footnote{\color{black}As can be seen in Fig.~\ref{fig:PLDPCH_DEC}, one control logic is responsible
for reading data from RAMs and loading them to the inputs of the Hadamard sub-decoders
while the other control logic is responsible
for writing the outputs of the Hadamard sub-decoders
to the RAMs. 
These two control logics can be combined into one, but are shown as two for clarity.}.


Using the read/write operations described in Section \ref{sect:wr_rams}, each set of LLRs, i.e., $z_2$ LLRs or $z_2$ vectors, is first divided into $G$ groups and then each group of LLRs
is stored in one of the $N_h$ RAMs, where $G = z_2 / N_h$.
With this storage method, we can retrieve $N_h$ values of $\{{L}_{app}^{\rm PVN}(\beta)\}$ (or $\{{L}_{ch}^{\rm PVN}(\beta)\}$), $N_h$ values of $\{{L}_{ex}^{\rm H}(\alpha,\beta)\}$ and $N_h$ vectors of $\{{\bm L}_{ch}^{\rm D1H(\alpha)}\}$ from the $N_h$ RAMs
in each clock cycle when single-port RAMs are used; and twice the number of LLRs values/vectors
when dual-port RAMs are used.
Once $d N_h$ values of $\{{L}_{app}^{\rm PVN}(\beta)\}$ (or $\{{L}_{ch}^{\rm PVN}(\beta)\}$), $d N_h$ values of $\{{L}_{ex}^{\rm H}(\alpha,\beta)\}$ and $d N_h$ vectors of $\{{\bm L}_{ch}^{\rm D1H(\alpha)}\}$
are retrieved,  the $N_h$ Hadamard sub-decoders
can operate on these $N_h$ individual batches of independent data.
To ensure that no conflict of memory access occurs during the decoding process, we design the size and storage of RAMs as follows.
\begin{itemize}
\item $N_h$ RAMs, denoted by PVN-CH-RAM,
are used to store $\{{L}_{ch}^{\rm PVN}(\beta):\beta=0,1,\ldots,N-1\}$.
Each RAM has a width of $w_{ch}^{\rm PVN}$ bits (to represent the quantized LLR value) and a depth of $n z_1 G$.
Referring to Fig. \ref{fig:RAM}, the $g$-th location ($g=0,1,\ldots,n z_1G-1$) in the $l$-th RAM
($l=0,1,\ldots,N_h-1$) stores ${L}_{ch}^{\rm PVN}(\beta)$ where $\beta=\lfloor g/G \rfloor z_2 + lG + (g\mod G)$.
Note that $\{{L}_{ch}^{\rm PVN}(\beta)\}$ is needed only once during the first decoding iteration.
After the first iteration, the content in PVN-CH-RAM is overwritten by the incoming channel LLR values of the next codeword.

\item
$N_h$ RAMs, denoted by PVN-APP-RAM,
are used to store $\{{L}_{app}^{\rm PVN}(\beta):\beta=0,1,\ldots,N-1\}$.
Each RAM has a width of $w_{app}^{\rm PVN}$ bits
 and a depth of $n z_1 G$.
Data are stored in the same way as in PVN-CH-RAM, i.e.,
the $g$-th location ($g=0,1,\ldots,n z_1 G - 1$) in the $l$-th RAM
($l=0,1,\ldots,N_h - 1$) stores ${L}_{app}^{\rm PVN}(\beta)$ where {\color{black}$\beta=\lfloor g/G \rfloor z_2 + lG + (g\mod G)$}.

\item
$N_h$ RAMs, denoted by H-EX-RAM,
are used to store $\{{L}_{ex}^{\rm H}(\alpha,\beta):\alpha=0,1,\ldots,M-1;\beta\in \{\beta_0, \beta_1, \ldots, \beta_{d-1}\} = \mathcal{P}(\alpha) \}$.
Each RAM has a width of $w_{ex}^{\rm H}$ bits
 and a depth of $m d z_1 G$.
Referring to Fig. \ref{fig:H-RAM}(a), the $q$-th location ($q=0,1,\ldots,m d z_1 G - 1$) in the $l$-th RAM
($l=0,1,\ldots,N_h - 1$) stores $\{{L}_{ex}^{\rm H}(\alpha,\beta)\}$ where
$\alpha=\lfloor q/d \rfloor N_h + l$, $\beta= \beta_{\delta}$, 
 and $\delta=q \mod d$.

\item
$N_h$ RAMs, denoted by D1H-CH-RAM,
are used to store $\{{\bm L}_{ch}^{\rm D1H(\alpha)}:{\color{black}\alpha=0,1,\ldots,M-1}\}$.
Each RAM has a width of $w_{ch}^{\rm D1H} = w_{ch}^{\rm PVN} \times (2^r-r-2)$ bits
 and a depth of $m z_1 G$.
Each address stores all the $2^r-r-2$ channel LLR values for D1H-VNs connected to a H-CN.
Referring to Fig. \ref{fig:H-RAM}(b), the $w$-th location ($w=0,1,\ldots,m z_1 G-1$) in the $l$-th RAM
($l=0,1,\ldots,N_h-1$) stores $\{{\bm L}_{ch}^{\rm D1H(\alpha)}\}$ where $\alpha = w N_h + l$.
(To allow the decoding to proceed while receiving the incoming channel LLR values of
the next codeword, either two sets of D1H-CH-RAM are used or
the depth of D1H-CH-RAM is doubled to
$2 m z_1 G$. We double the depth of D1H-CH-RAM to
$2 m z_1 G$ in our design.)
{\color{black}Moreover, we use dual-port RAMS --- one port reads the data in D1H-CH-RAM used for decoding
and the other port writes  incoming channel LLR values into the same RAM.}
\end{itemize}

\subsection{Latency and Throughput}\label{subsect:l&t}

\subsubsection{$N_h = z_2$} \label{sect:N_h=z_2}

We first consider a special case
in which maximum parallelism is designed for each layer.
In other words, we consider the case where $N_h = z_2$ and $G = z_2 / N_h = 1$.
We also assume dual-port RAMs are used and hence two memory addresses can be accessed at the same
time \footnote{\color{black} Note that single-port RAMs can also be used but then only one memory address can be accessed at one
time. The derivations of latency and throughput would be similar to those described in this section but the results would be worse.} and it takes $d/2$ clock cycles to retrieve
the required $d$ sets of ${L}_{app}^{\rm PVN}(\beta)$ (or ${L}_{ch}^{\rm PVN}(\beta)$) and ${L}_{ex}^{\rm H}(\alpha,\beta)$ values in each layer.
Note that  $\{{L}_{ex}^{\rm PVN}(\alpha,\beta)\}$ in \eqref{eq:cpt_LexPVN2} is computed
in the same clock cycle
 as ${L}_{app}^{\rm PVN}(\beta)$ (or ${L}_{ch}^{\rm PVN}(\beta)$) and ${L}_{ex}^{\rm H}(\alpha,\beta)$ are retrieved.
At the $d/2$-th clock cycle, we also load the required $z_2$ sets of $\bm{L}_{ch}^{\rm{D1H}(\alpha)}$
from one address location to the sub-decoders.
Subsequently, $dz_2$ LLRs of $\{{L}_{ex}^{\rm PVN}(\alpha,\beta)\}$ and $z_2$ vectors of  $\{\bm{L}_{ch}^{\rm{D1H}(\alpha)}\}$
are passed to the $z_2$ FHT blocks in the $z_2$ Hadamard sub-decoders,
i.e., $d$ LLRs of $\{{L}_{ex}^{\rm PVN}(\alpha,\beta)\}$ and one vector of  $\{\bm{L}_{ch}^{\rm{D1H}(\alpha)}\}$
 to one FHT block in one Hadamard sub-decoder.
Then, it takes $2r+1$ clock cycles to compute $dz_2$ LLRs of
$\{{L}_{ex}^{\rm H}(\alpha,\beta)\}$ and $dz_2$ LLRs of $\{{L}_{app}^{\rm PVN}(\beta)\}$ using \eqref{eq:L_ex_H_layered}
and  \eqref{eq:L_app_PVN_layered}, respectively.
Finally, it takes another $d/2$ clock cycles to write
these updated ${L}_{app}^{\rm PVN}(\beta)$ and ${L}_{ex}^{\rm H}(\alpha,\beta)$ values into the RAMs.

To summarize,
\begin{enumerate}
\renewcommand{\labelenumi}{\roman{enumi})}
\item Clock cycle no. $1$ to $d/2$: read $\{{L}_{app}^{\rm PVN}(\beta)\}$ (or ${L}_{ch}^{\rm PVN}(\beta)$) and $\{{L}_{ex}^{\rm H}(\alpha,\beta)\}$ from memory, {\color{black} and at the same time compute $\{{L}_{ex}^{\rm PVN}(\alpha,\beta)\}$ using \eqref{eq:cpt_LexPVN2}};
\item Clock cycle no. $d/2$ (in parallel with above): read $\{\bm{L}_{ch}^{\rm{D1H}(\alpha)}\}$;
\item Clock cycle no. $d/2+1$ to $d/2+2r$: process the inputs $\{{L}_{ex}^{\rm PVN}(\alpha,\beta)\}$ and $\{\bm{L}_{ch}^{\rm{D1H}(\alpha)}\}$  by the Hadamard sub-decoders (consisting of FHT and DFHT blocks) using \eqref{eq:cpt_LappH};
\item Clock cycle no. $d/2+2r+1$: compute $\{{L}_{ex}^{\rm H}(\alpha,\beta)\}$ and $\{{L}_{app}^{\rm PVN}(\beta)\}$ {\color{black}using \eqref{eq:L_ex_H_layered} and \eqref{eq:L_app_PVN_layered}};
\item Clock cycle no. $d/2+2r+2$ to $d/2+2r+1+d/2$: write $\{{L}_{app}^{\rm PVN}(\beta)\}$ and $\{{L}_{ex}^{\rm H}(\alpha,\beta)\}$ to memory.
\end{enumerate}
Since $d=r+2$, the whole process takes $d/2+2r+1+d/2=3r+3$ clock cycles.

When $N_h = z_2$,  maximum parallelism for each layer is achieved.
The latency is minimized and  the throughput of the decoder is maximized.
However, such a design consumes a lot of hardware resources (a large number of RAMs and
$N_h$ Hadamard sub-decoders) and may not be practical.
In the next section, we consider the cases when $N_h$ is smaller than $z_2$.

\subsubsection{$N_h < z_2$}
We consider the case when $N_h < z_2$ and $G=z_2/N_h$ is an integer.
Using the proposed decoder architecture,
$G(>1)$ groups of H-CNs (each consisting of $N_h$ H-CNs) are {\color{black}sequentially} processed in each layer.
Referring to the timing details in Section~\ref{subsec:str_h}  and Section~\ref{sect:N_h=z_2} and with the use of
our RAM designs,
it takes $d/2$ clock cycles to load the data of one group of H-CNs.
(Recall that dual-port RAMs are used.)
We use a pipelined structure and load the $G$ groups of
data to the sub-decoders in  a consecutive manner.
To complete loading all $G$ groups of data, it takes $t_{loading}=dG/2$ clock cycles.
Moreover, the first set of outputs (i.e., ${L}_{app}^{\rm PVN}(\beta)$ and ${L}_{ex}^{\rm H}(\alpha,\beta)$)
 is available at the $t_{1st\;output}=(d/2+2r+1)$-th clock cycle.

%

 \begin{figure*}[t]
\centerline{
\includegraphics[width=1\columnwidth]{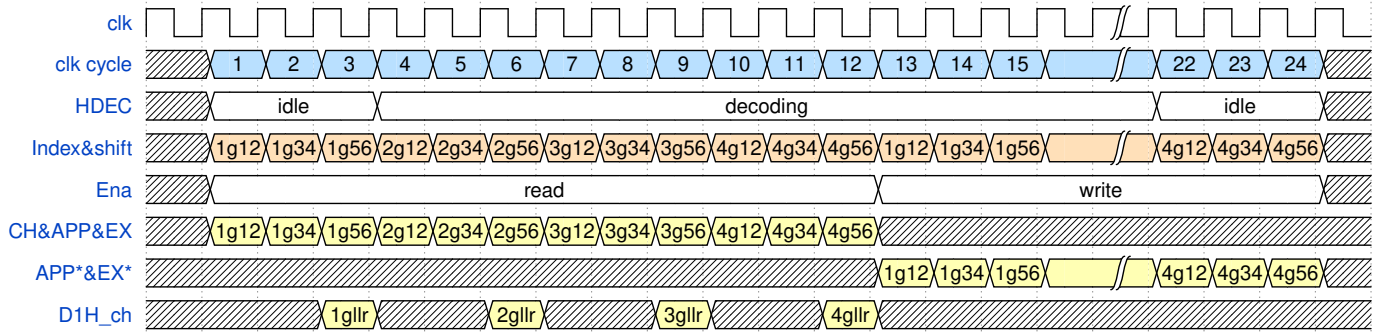}}
 \caption{Timing diagram for the decoding of one layer of PLDPC-Hadamard code.  $r=4$, $z_2 = 512$, $N_h = 128$ and $G = z_2 / N_h = 4$. HDEC represents the state of the Hadamard sub-decoder; Index$\&$shift represents the shift values of the corresponding CPMs; Ena represents
 the state of PVN-CH-RAMs, PVN-APP-RAMs and H-EX-RAMs; CH$\&$APP$\&$EX represent the channel LLR values of $\{{L}_{ch}^{\rm PVN}(\beta)\}$, the \textit{a posteriori} LLRs of $\{{L}_{app}^{\rm PVN}(\beta)\}$ and the extrinsic LLRs of $\{{L}_{ex}^{\rm H}(\alpha,\beta)\}$; APP*$\&$EX* represents the updated LLRs for $\{{L}_{app}^{\rm H}(\beta)\}$ and the updated extrinsic LLRs for $\{{L}_{ex}^{\rm H}(\alpha,\beta)\}$; D1H$\_$ch represents the channel LLRs of $\{{\bm L}_{ch}^{\rm D1H(\alpha)}\}$. }
  \label{fig:Nh128}
\end{figure*}

\begin{figure*}[t]
\centerline{
\includegraphics[width=1\columnwidth]{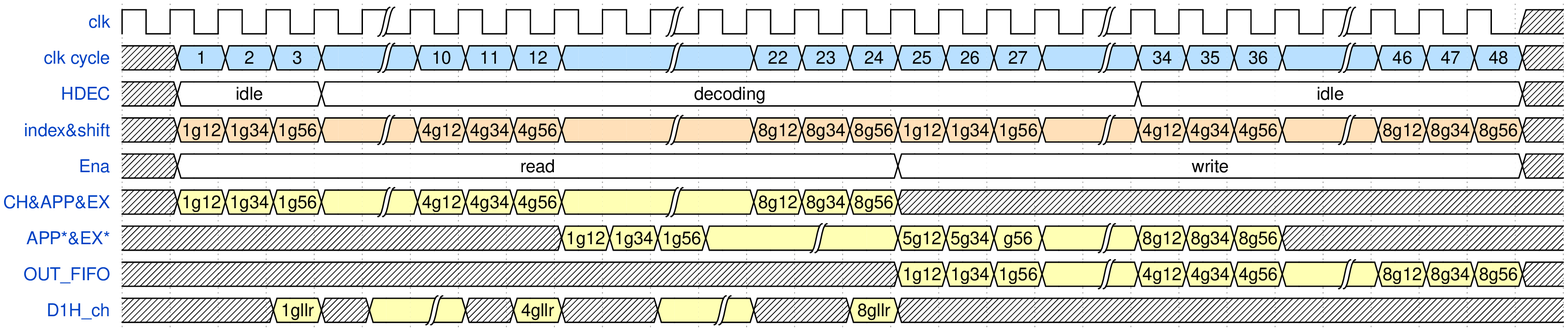}}
 \caption{Timing diagram for the decoding of one layer of PLDPC-Hadamard code. $r=4$,  $z_2 = 512$,
 $G=8$ and $N_h = 64$. OUT$\_$FIFO represents the output LLRs for $\{{L}_{app}^{\rm H}(\beta)\}$ and $\{{L}_{ex}^{\rm H}(\alpha,\beta)\}$. The representations of other symbols  are the same as in Fig. \ref{fig:Nh128}. }
  \label{fig:Nh64}
\end{figure*}

\paragraph{Case I $t_{loading} \leq t_{1st\;output}$} \label{sect:case-I}

It means that all the required data are read from the RAMs before the Hadamard sub-decoders generate the updated results.
The total time taken to complete updating one layer equals
``loading time of all groups + processing time of last group + writing time of last group'', i.e.,
\begin{eqnarray}
t_{l1} &=& t_{loading} + (2r+1) + d/2 \nonumber\\
&=& (r/2 + 1)G + 5r/2+2
\label{eq:total_time_1}
\end{eqnarray}
using $d=r+2$.
Supposing $I$ iterations are needed and the clock frequency is $f_c$, the latency for decoding each codeword equals
\begin{eqnarray}
t_{c1} &=& I m z_1 t_{l1} /f_c  \nonumber\\
&=& I m z_1 [(r/2 + 1)G  + 5r/2+2] /f_c,  
\end{eqnarray}
where $mz_1$ is the number of layers in layered decoding.
For a given $m \times n$ base matrix, the latency $t_{c1}$ can be reduced by (a) lowering $I$ and/or $z_1$ and/or $G$; or (b) increasing $f_c$.
As the codeword length is $l={n{z_1}{z_2} + m{z_1}{z_2}\left( {{2^r} - r - 2} \right)}$,
the throughput of the decoder is expressed as
\begin{eqnarray}
T_1 &=& \frac{l}{t_{c1}} = \frac{{\left[ {n{z_1}{z_2} + m{z_1}{z_2}\left( {{2^r} - r - 2} \right)} \right]  {f_c}}}{I m z_1 t_{l1} } \nonumber\\
 & {\color{black}=} & \frac{{\left[ {n/m + \left( {{2^r} - r - 2} \right)} \right] z_2 {f_c}}}{I   [ (r/2 + 1)G  + (5r/2+2) ] }.
\end{eqnarray}
To improve the throughput, we can
(a) increase $z_2$  and/or $f_c$;
or (b) decrease $I$ and/or $G$.

{\underline{Example}}:
Taking $r=4$, $d = r + 2 = 6$, $z_2 = 512$ and $N_h = 128$ as an example,
we have $t_{loading}=dG/2=12$ and $t_{1st\;output}=(d/2+2r+1)=12$.
 Fig. \ref{fig:Nh128} shows the  timing diagram  for the decoding of one layer, in which
 the LLR data is divided into $G = z_2 / N_h = 4$ groups.
\begin{enumerate}
\renewcommand{\labelenumi}{\roman{enumi})}
\item {Clock cycle no. $1$ to $dG/2 = 12$:}
We load $G = 4$ groups of $\{{L}_{ex}^{\rm PVN}(\alpha,\beta)\}$ into
the Hadamard sub-decoders corresponding to the $z_2$ H-CNs in the layer in $dG/2 = 12$ clock cycles.
In each clock cycle,
 $2 N_h= 256$ LLRs of $\{L_{app}^{\rm PVN}(\beta)\}$ (or ${L}_{ch}^{\rm PVN}(\beta)$) and $256$ LLRs of  $\{{L}_{ex}^{\rm H}(\alpha,\beta)\}$ are read from RAMs, and at the same time  $256$  LLRs of $\{{L}_{ex}^{\rm PVN}(\alpha,\beta)\}$ are computed using \eqref{eq:cpt_LexPVN2} and loaded into
 the $N_h= 128$ Hadamard sub-decoders.
 Therefore, it takes $d/2=3$ clock cycles to completely retrieve all LLR values belonging
 to the first group, i.e.,  $d N_h= 6 \times 128 = 768$ LLRs of $\{L_{app}^{\rm PVN}(\beta)\}$ (or ${L}_{ch}^{\rm PVN}(\beta)$) and $768$ LLRs  of $\{{L}_{ex}^{\rm H}(\alpha,\beta)\}$, and to compute and load $768$  LLRs of $\{{L}_{ex}^{\rm PVN}(\alpha,\beta)\}$ into
 the $N_h= 128$ Hadamard sub-decoders.
Referring to Fig. \ref{fig:Nh128}, we use the symbol
``1g12" to represent the LLRs corresponding to the first and second  P-VNs in Group \#1,
``1g34" to represent the LLRs corresponding to the third and fourth P-VNs  in Group \#1, and
``1g56" to represent the LLRs corresponding to the fifth and sixth P-VNs  in Group \#1.
Moreover, ``$Z$g12",  ``$Z$g34" and ``$Z$g56" where $Z=2,3,4$ are defined in a similar fashion.
Thus,  LLR values belonging to Group  \#1 are retrieved during Clock cycle no. \#1 to \#3;
Group \#2 during Clock cycle no. \#4 to \#6;
Group \#3 during Clock cycle no. \#7 to \#9; and
Group \#4 during Clock cycle no. \#10 to \#12.
%
%
%
%
\item
Clock cycle no. $3$, $6$, $9$ and $12$: We load $G = 4$ groups of $\{\bm{L}_{ch}^{\rm{D1H}(\alpha)}\}$ into the Hadamard sub-decoders.
At clock no. $3$, we load the channel LLRs for D1H-VNs in Group \#1
 into
 the $N_h= 128$ Hadamard sub-decoders.
Referring to ``D1H$\_$ch'' in Fig. \ref{fig:Nh128}, we use the symbol ``1gllr" to represent
these LLRs in Group \#1.
Similarly, at clock no. $6$, $9$ and $12$, we load the channel LLRs for D1H-VNs in Group \#2,
Group \#3 and Group \#4, respectively,  into
 the $N_h= 128$ Hadamard sub-decoders.
 They are represented by  ``$Z$gllr"  in Fig. \ref{fig:Nh128} where $Z=2,3,4$
 \footnote{\color{black} Note that here for convenience, we use only $1$ clock cycle to load one group of  $\{\bm{L}_{ch}^{\rm{D1H}(\alpha)}\}$ into the Hadamard sub-decoders. 
Thus the four groups of $\{\bm{L}_{ch}^{\rm{D1H}(\alpha)}\}$ are loaded during clock cycle nos. $3$, $6$, $9$ and $12$, leaving some ``blank regions''
 between these clock cycles in Fig. \ref{fig:Nh128} (and also in Fig. \ref{fig:Nh64}). 
 Another design is to load each group of $\{\bm{L}_{ch}^{\rm{D1H}(\alpha)}\}$ using multiple clock cycles (a maximum of 3 clock cycles in this example) so as to minimize
 the  span of the ``blank regions''.}.
%
%
%
\item Clock cycle no. $4$ to $21$:
We decode one layer
consisting of $z_2$ H-CNs in a pipeline manner.
At Clock cycle no. $4$, the Hadamard sub-decoders starts processing the LLRs belonging to Group  \#1
which has completed its LLR loading at Clock cycle no. $3$.
Similarly, at Clock cycle no. $7$, $10$ and $13$, the Hadamard sub-decoders starts processing the LLRs belonging to  Group \#2,
Group \#3 and Group \#4, respectively.
Since it takes $2r+1=9$ clock cycles to process each group of LLRs
and the groups of LLRs are processed in a pipeline manner,
the last group of LLRs will be processed completely at Clock cycle no. $13+9-1=21$.
\item
Clock cycle no. $13$ to $t_{l1} = 24$: We write the updated $G = 4$ groups of data into the corresponding RAMs.
Referring to the step above, at Clock cycle no. $4$, $7$, $10$ and $13$, the Hadamard sub-decoders starts processing the LLRs belonging to Group  \#1,  Group \#2,
Group \#3 and Group \#4, respectively. Moreover,
at Clock cycle no. $12$, $15$, $18$ and $21$, the Hadamard sub-decoders has completed processing the LLRs belonging to Group  \#1,  Group \#2,
Group \#3 and Group \#4, respectively;
and has each time generated a group of LLRs consisting of
$768$ LLR values of $\{{L}_{app}^{\rm PVN}(\beta)\}$ and $768$ LLR values of $\{{L}_{ex}^{\rm H}(\alpha,\beta)\}$.
In a similar fashion as in Step i), it takes $d/2$ clock cycles to store/write the LLRs belonging to one group. Thus, LLRs belonging to Group  \#1 are stored during Clock cycle no. $13$ to $15$;
Group  \#2 stored during Clock cycle no. $16$ to $18$;
Group  \#3 stored during Clock cycle no. $19$ to $21$;
Group  \#4 stored during Clock cycle no. $22$ to $24$.
In other words, from Clock cycle no. $13$ to $24$,  $2 \times 128 = 256$ updated LLR values of $\{{L}_{app}^{\rm PVN}(\beta)\}$ are stored into $N_h=128$ PVN-APP-RAMs and $2 \times 128 = 256$ updated LLR values of $\{{L}_{ex}^{\rm H}(\alpha,\beta)\}$ are stored to $N_h=128$ H-EX-RAMs during each clock cycle.

Note that the total time taken to complete updating one layer is $24$ clock cycles, which
is the same as the theoretical result computed using \eqref{eq:total_time_1}.
\end{enumerate}

%
%
%

\paragraph{Case II $t_{loading} > t_{1st\;output}$}
It means that the Hadamard sub-decoders start to output the updated results before
all the required data have been read from the RAMs.
In this case, we need to use first-in-first-out (FIFO) RAMs to temporarily store the updated results
(i.e., ${L}_{app}^{\rm PVN}(\beta)$ and ${L}_{ex}^{\rm H}(\alpha,\beta)$) from
the Hadamard sub-decoders. Once all the required data are read from the RAMs,
the updated results stored in the FIFO RAMs are written to the RAMs.
The total time taken to complete updating one layer equals
``loading time of all groups + writing time of all groups'', i.e.,
\begin{equation}
t_{l2} = dG/2 + dG/2  = (r + 2)G.
\end{equation}
The latency to decode one codeword equals
\begin{eqnarray}
t_{c2} = I m z_1 G (r + 2) /f_c,  
\end{eqnarray}
and the throughput 
equals
\begin{eqnarray}
T_2 = \frac{{\left[ {n/m + \left( {{2^r} - r - 2} \right)} \right]  {f_c}{z_2}}}{I  G(r + 2) }
\end{eqnarray}
which can be improved by
(a) increasing $f_c$ and/or $z_2$;
or (b) decreasing $I$ and/or $G$.
Fig.~\ref{fig:Nh64} shows the timing diagram when decoding one layer with parameters $z_2 = 512$,
$N_h = 64$ and $G= z_2 / N_h = 8$.
The difference between this case and  the previous one is that we use FIFO RAMs to temporarily store the ``updated" LLR values until
all the required data are loaded into Hadamard sub-decoders.

Note that in both Case I and Case II, it requires $d/2$ clock cycles to complete loading one group of data
into the $N_h$ Hadamard sub-decoders.
Thus, the $N_h$ Hadamard sub-decoders are idle most of the time \footnote{\color{black}Note that it is possible to 
start processing the next layer before the current one is entirely completed.
Memory access conflicts as in conventional QC-LDPC decoding architectures will occur but can also be resolved by methods introduced in Sect.~\ref{sect:Introduction}.
As this paper mainly  focuses on realizing the PLDPC-HC layered decoder and estimates its fixed-point error performance, improving the throughput of the decoder would be left for our future work.}.
Therefore the throughput can potentially be increased by a factor of $d/2$
if the Hadamard sub-decoders are allowed to process $d/2$ different codewords at the same time.
The extra requirement would be $d/2$ times increase in memory storage and a bit more control logics \cite{Kumawat2015}.

%

\section{Implementation Results }\label{sect:results}
We implement the $r=4$ and $R=0.0494$ PLDPC-Hadamard decoder (whose base matrix and protograph are shown in Fig. \ref{fig:base&proto}) optimized in \cite{zhang2020proto,zhang2021} on the  Xilinx VCU118 FPGA board.
The maximum operating frequency is $f_c=130$~MHz \footnote{\color{black}We start the experiments by setting the clock frequency of the FPGA to  $80$ MHz.
After confirming that the experimental results (i.e., error rates) are the same as those given
by fixed-point computer simulations, we continually increase the clock frequency of the FPGA. When the clock  frequency exceeds $130$ MHz, the experimental results are no longer the  same as those given
by fixed-point computer simulations. Thus we claim a maximum operating frequency 
of  $f_c=130$ MHz.} {\color{black}and true dual-port RAMs are used}.
Binary phase-shift-keying (BPSK) modulation
and an additive white Gaussian noise channel are assumed.
To compare with the floating-point results in {\color{black}\cite{Zhang2021layer}}, we use the same lifting factors, i.e.,
 $z_1 = 32$ and $z_2 = 512$, and the same code length $l=1327104$.

We implement two designs with $N_h = 128$ ($G=4$) and $N_h = 64$ ($G=8$) Hadamard sub-decoders, respectively, which belong to Case I and Case II in Section \ref{subsect:l&t}.
{\color{black}First, we consider the bit-widths setting $S1$ shown in Fig. \ref{fig:data_flow}(a) that has been
 implemented for both designs.   
Fig. \ref{fig:fixed_S1S2S3} plots the  FER/BER results of the PLDPC-Hadamard code
when the number of iterations $I=20$ and $150$.}
It can be observed that the two designs (i.e., $N_h = 128$ and $N_h = 64$) 
with bit-widths setting $S1$ produce almost the same FER/BER curves. The
minute difference arises only because the same noise samples generated have been assigned to
different code bits in the two different designs.
The results also show that at a BER of $10^{-5}$,
the fixed-point decoder with bit-widths setting $S1$ suffers from a degradation of
$0.08$ dB compared with the floating-point computation when $I = 150$; and a degradation of $0.10$ dB when $I = 20$.
{\color{black}While no FER/BER error floors appear  for the floating-point simulations; for fixed-point results, error floors start to emerge (i) at a BER of
$3 \times 10^{-6}$ (FER around $1.05 \times 10^{-2}$) for $I = 150$ iterations  and 
(ii) at a BER of $10^{-6}$  (FER around $1.05 \times 10^{-2}$) for $I = 20$ iterations.

To investigate the effect of bit-widths setting on the error performance of the fixed-point decoder, 
we increase the integer part for all the types of LLRs (except for channel observations) in setting $S1$ by one bit and form 
the bit-widths setting $S2$ shown in Fig. \ref{fig:data_flow}(b).
We implement the setting $S2$ for the design with $N_h = 128$ Hadamard sub-decoders and plot the FER/BER results in Fig. \ref{fig:fixed_S1S2S3} with $I = 20$ iterations.  We can observe that
increasing the bit-widths can effectively remove 
 the FER/BER error floors when $I = 20$ iterations.
{ \color{black}Based on the setting $S2$, we increase the fractional part of (i) output of FHT, (ii) input of DFHT, and (iii) 4 (internal) stages in DFHT, by one bit and form the bit-widths setting $S3$. 
 In other words, all the above three categories are represented by 11 bits, i.e., 
change  from  ``1 sign + 7 int + 2 frac'' to ``1 sign + 7 int + 3 frac''. 
 We implement the setting $S3$ for the design with $N_h = 128$ Hadamard sub-decoders and plot the FER/BER results in Fig. \ref{fig:fixed_S1S2S3} with $I = 150$ iterations.
 Comparing the results for setting $S3$ and those for setting $S1$
 shows that 
no BER error floor is observed down to $2 \times 10^{-8}$  (for setting $S3$)
and the FER error floor is lowered from $10^{-2}$ 
(for setting $S1$) to $5 \times 10^{-5}$ (for setting $S3$). 
 }

\begin{figure*}[t]
\centerline{
\includegraphics[width=0.5\columnwidth]{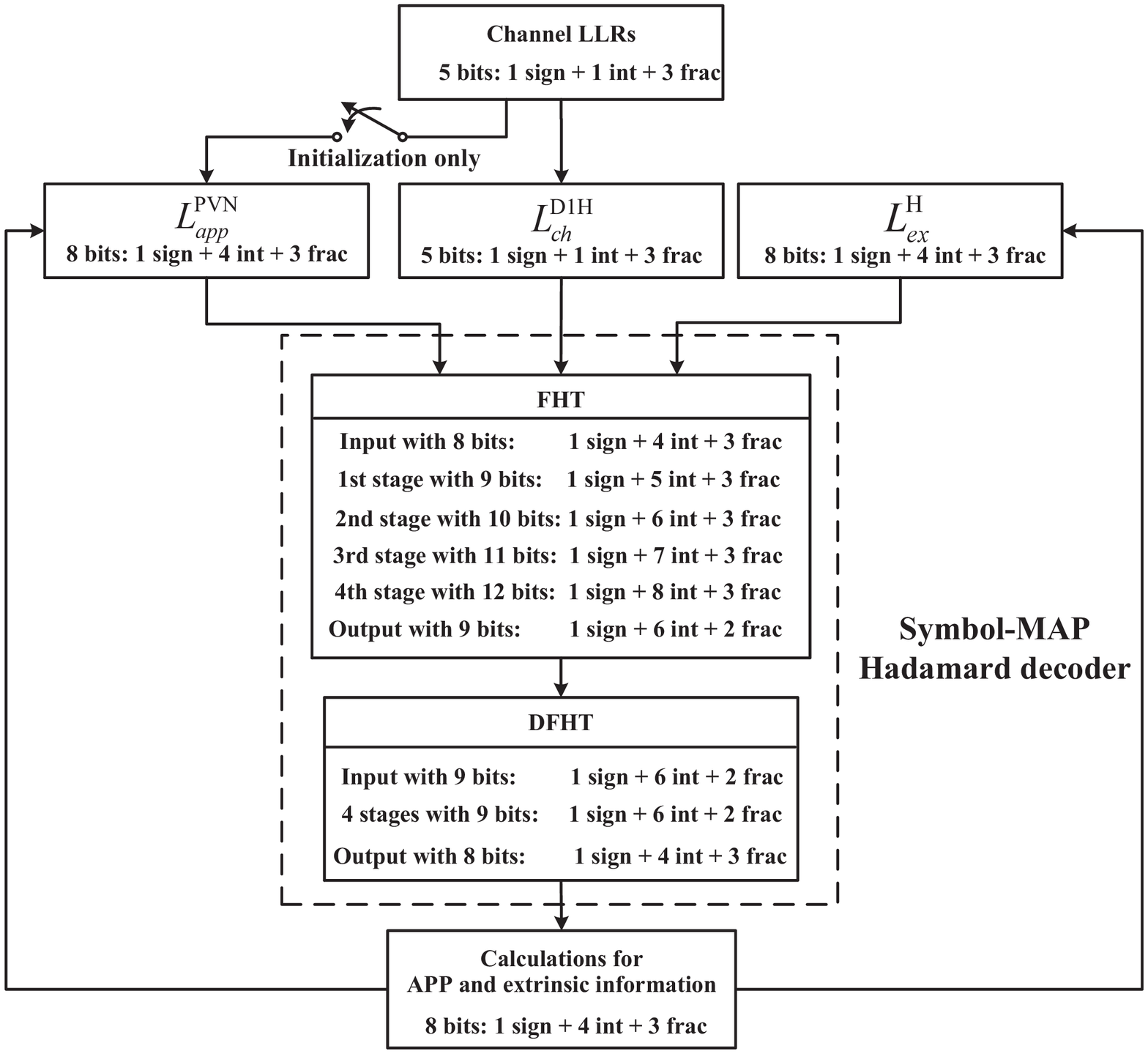}\;\;\;\;\;
\includegraphics[width=0.5\columnwidth]{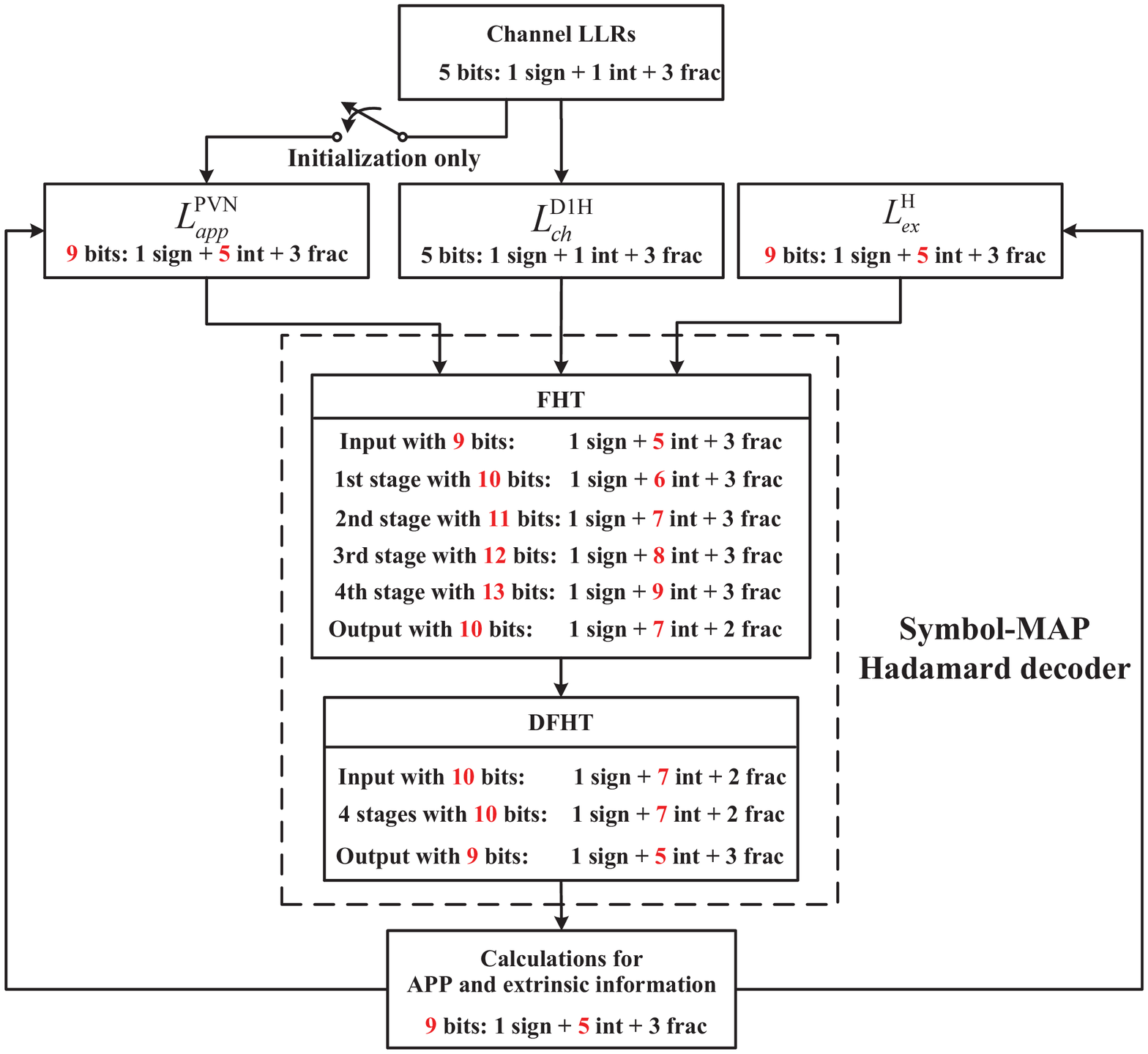}}
 \centerline{(a) \rule{8.8cm}{0cm} (b)}
 \caption{\color{black} Data transformation among different modules for a $r=4$ PLDPC-Hadamard code.
 ``$1$ sign + $y$ int + $z$ frac'' denotes $1$ bit to represent sign, $y$ bits to represent the integral part,
 and $z$ bits to represent the fractional part.
 (a) Bit-widths setting $S1$.
 (b) Bit-widths setting $S2$. 
\color{black}When the fractional parts of (i) output of FHT, (ii) input of DFHT, and (iii) 4 (internal) stages in DFHT, are increased by one bit, i.e., 
change from  ``1 sign + 7 int + 2 frac'' to ``1 sign + 7 int + 3 frac'', bit-widths setting $S3$ is formed.}
  \label{fig:data_flow}
\end{figure*}

\begin{figure*}[t]
\centerline{
\includegraphics[width=1\columnwidth]{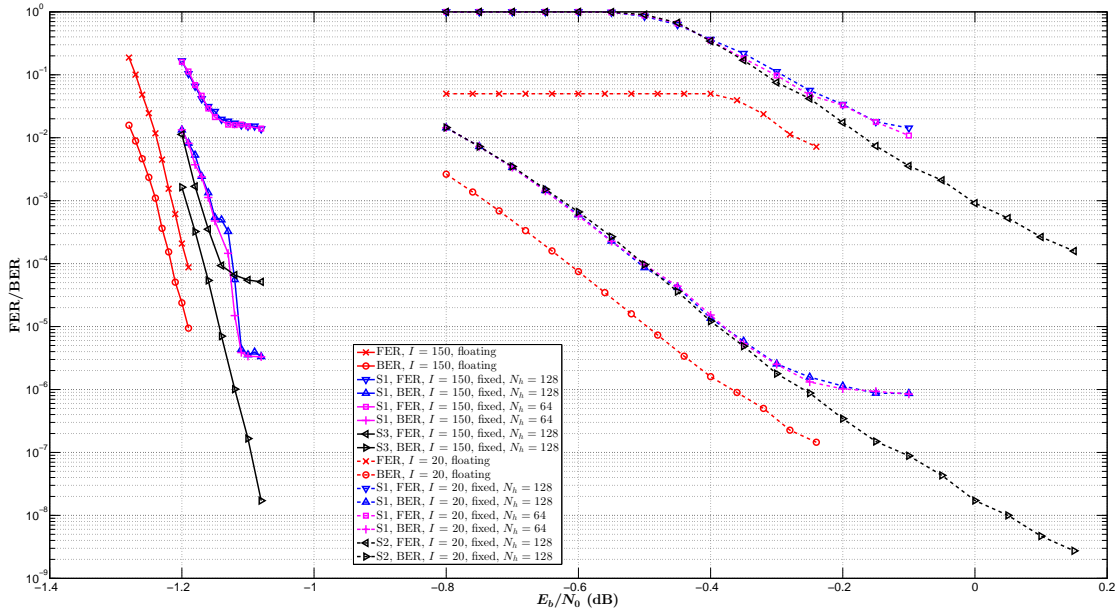}}
 \caption{\color{black}Floating-point and fixed-point BER/FER performance of the layered PLDPC-Hadamard decoders.
 Floating-point results are obtained by computer simulations and the fixed-point results are generated via the FPGA platform with bit-widths setting $S1$, $S2$ or $S3$.
$r = 4$,  $l = 1327104$, $I=20,150$ and  $N_h=64$ or $128$.}
  \label{fig:fixed_S1S2S3}
\end{figure*}

{\color{black}Finally, we consider the latency and hardware implementation of the decoders.}
 For $r=4$ (hence $d=r+2=6$), $t_{1st\;output}=(d/2+2r+1)=12$ cycles.
When $G = 4$, $t_{loading}=dG/2=12=t_{1st\;output}$ which belongs to Case I in Section \ref{subsect:l&t}. The decoding latency per layer equals $t_{l1}  = 24$ cycles. \footnote{In practice, there is a fixed delay $t_\delta$ when operating RAMs.
In our designs, $t_\delta=2$ cycles and are included in deriving the latency and throughput in Table \ref{tb:compare}.}
Similarly when $G = 8$, $t_{loading}=dG/2=24>t_{1st\;output}$ which belongs to Case II. The decoding latency per layer equals $t_{l2}  = 48$ cycles.
Table \ref{tb:compare} lists the hardware implementation results of the proposed layered decoder
for $N_h=64$ ($G=8$) and $N_h=128$ ($G=4$).
\footnote{\color{black}The vast majority of the hardware resources are used in  (i) storing the LLRs (with BRAMs)
and (ii)  implementing the parallel Hadamard sub-decoders (with LUTs).
The usage of the resources for other purposes is relatively very small. } 
Since the code lengths are identical, the two designs
 consume almost the same amount of block RAMs (BRAMs).
Compared with the decoder with $N_h = 64$ Hadamard sub-decoders
{\color{black}and under the same bit widths setting $S1$,}
the one with $N_h = 128$ sub-decoders produces about
twice the throughput, reduces the latency by about half, and utilizes about twice the amount of look-up tables (LUTs).
{\color{black} 
For the decoder with $N_h = 128$ sub-decoders, increasing the 
bit widths from setting $S1$ to setting $S2$ increases the LUT utilization
from
$81.92\%$ to $93.29\%$ but does not change the amount of BRAMs
used. {\color{black}Note that the hardware utilization of bit widths setting $S3$
is almost the same as that of bit widths setting $S2$ and is therefore not shown in  Table \ref{tb:compare}.
}

\begin{table}[t]
\newcommand{\tabincell}[2]{\begin{tabular}{@{}#1@{}}#2\end{tabular}}
\caption{Comparison of implementation results for PLDPC-Hadamard decoder with $64$ and $128$ Hadamard sub-decoders. Bit-widths settings $S1$ and $S2$ are used. Hadamard order $r = 4$, code rate $R = 0.0494$, code length $l = 1327104$, and clock frequency $f_c = 130$ MHz.  LUT: Look-up Table; BRAM: Block RAM.}\label{tb:compare} 
\small
\begin{center}
\setlength{\tabcolsep}{0.7mm}{
\begin{tabular}{|c|c|c|c|c|}
\hline
{\tabincell{c}{Available \\ LUT  }} & \multicolumn{4}{|c|}{$1,182,240$ } \\
\hline
{\tabincell{c}{Available \\ BRAM }} & \multicolumn{4}{|c|}{$2,160$ } \\
\hline
\hline
{\tabincell{c}{No. of\\ sub-decoders}} & \multicolumn{2}{|c|}{$N_h=64$ } & \multicolumn{2}{|c|}{ $N_h=128$ }  \\
\hline
{\tabincell{c}{S1: LUT \\ Utilization }} & \multicolumn{2}{|c|}{$485,738$ ($41.09\%$) } & \multicolumn{2}{|c|}{ $968,538$ ($81.92\%$) }  \\
\hline
{\tabincell{c}{S1: BRAM \\ Utilization}} & \multicolumn{2}{|c|}{$718.5$ ($33.26\%$) } & \multicolumn{2}{|c|}{ $715$ ($33.10\%$) }  \\
\hline
{\tabincell{c}{S2: LUT \\ Utilization }} & \multicolumn{2}{|c|}{ NA } & \multicolumn{2}{|c|}{ $1,102,958$ ($93.29\%$) }  \\
\hline
{\tabincell{c}{S2: BRAM \\ Utilization}} & \multicolumn{2}{|c|}{ NA } & \multicolumn{2}{|c|}{ $715$ ($33.10\%$) }  \\
\hline
\hline
{\tabincell{c}{No. of\\ iterations}} &$I = 150$&  $I = 20$ & $I = 150$&  $I = 20$  \\
\hline
{\tabincell{c}{$E_b/N_0$ at \\  BER of $10^{-5}$ }} &$-1.11$ dB  &  $-0.40$ dB &$-1.11$ dB  &$-0.40$ dB   \\
\hline
Latency &$12.92$ ms &  $1.72$ ms  & $6.72$ ms &  $0.896$ ms  \\
\hline
{\tabincell{c}{\color{black}Coded \\ throughput}} & $0.10$ Gbps & $0.77$ Gbps  & $0.20$ Gbps &  $1.48$ Gbps  \\
\hline
\end{tabular}}
\end{center}
\end{table}

\section{Conclusion}\label{sect:conclusion}
A hardware architecture of the PLDPC-Hadamard layered decoder has been designed and implemented onto an FPGA.
The architecture consists of  control logics, BRAMs and Hadamard sub-decoders.
The latency and throughput of the design have been derived
in the terms of the code parameters and the amount of parallel sub-decoders deployed.
A throughput of $1.48$ Gbps is achieved when $20$ decoding iterations are used. 
{\color{black}Moreover, increasing the bit widths of the data in the decoder 
has shown to be an effective way of {\color{black}eliminating/lowering} the
FER/BER error floors.}  

In our current decoder design,
the Hadamard sub-decoders are not fully utilized
in the time domain. When these sub-decoders are fully utilized, the decoder
can decode $d/2$ ($=3$ in the example used) codewords simultaneously
and hence increase the throughput by the same factor (i.e., to almost $4.5$ Gbps).
To decode more codewords, more BRAMs would be needed though.
Our decoder architecture is generic and can be readily modified to decode LDPC-Hadamard codes with the order of the Hadamard code being odd, i.e., $r$ is odd.
Moreover, it can be modified and applied to
decode other LDPC-derived codes when the Hadamard constraints LDPC-HC are replaced by other code constraints.

{\color{black}To {\color{black}eliminate/lower} the error floors, we have increased the bit widths in our decoder design.
Another future direction of research is to design variable bit widths (i.e., quantization schemes) in
the decoder \cite{ZhangX2014} so as to reduce the complexity of the decoder and 
to remove the error floor issue.}
 


\bibliographystyle{ieeetr}
\bibliography{Reference}

\begin{IEEEbiography}
[{\includegraphics[width=1in,height=1.25in,clip,keepaspectratio]{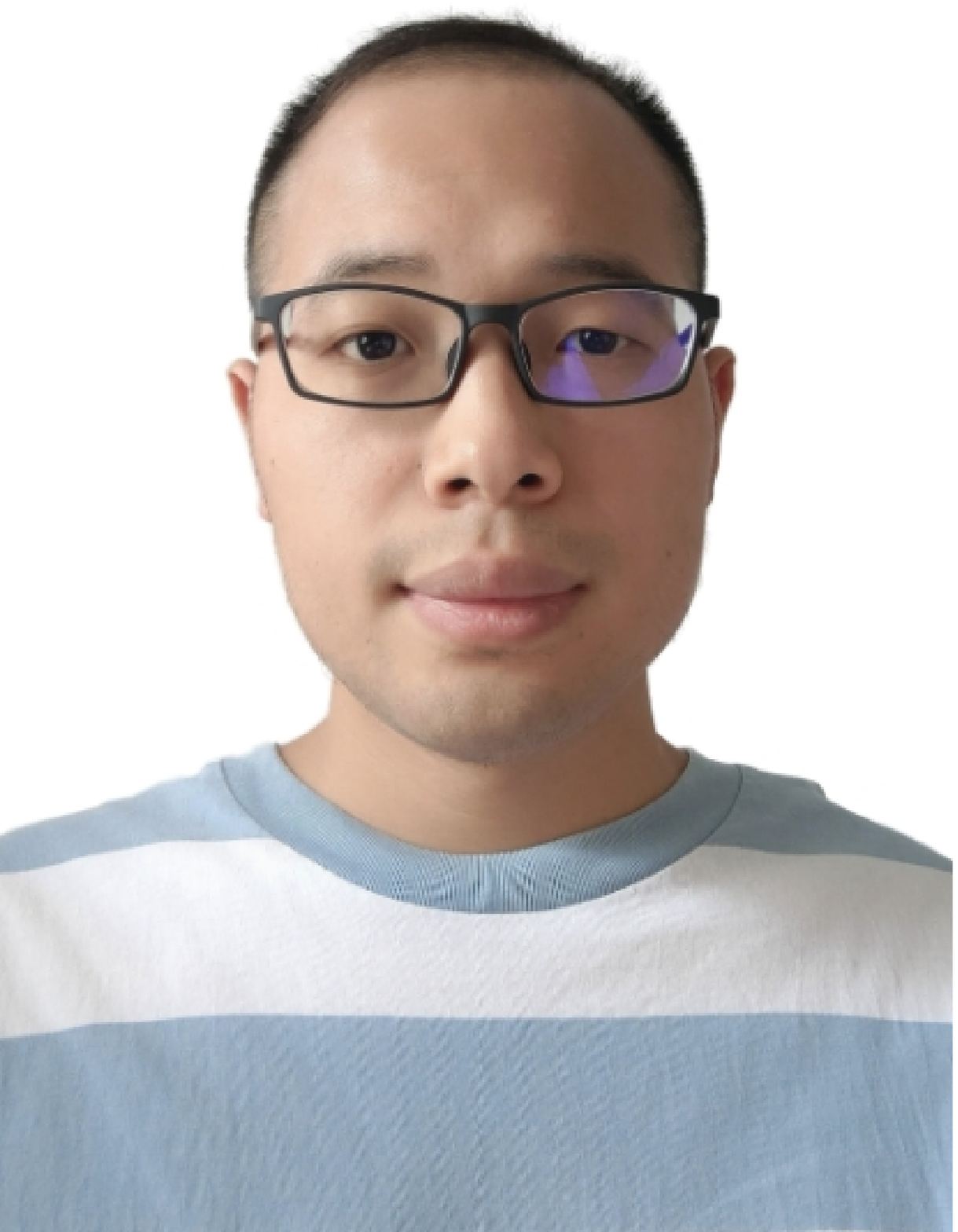}}]
{Peng-Wei Zhang} received the Bachelor of Engineering degree in Electronics and Information Engineering and
the Master of Engineering degree in Electronics and Communication Engineering from Chongqing University of Posts and Telecommunications, China, in 2013 and 2016, respectively.
He received his PhD degree at the Department of Electronic and Information Engineering, The Hong Kong Polytechnic University, Hong Kong SAR, China, in 2021. He is now with Huawei Technologies Ltd., Chengdu, China.
\end{IEEEbiography}

\begin{IEEEbiography}[{\includegraphics[width=1in,height=1.25in,clip,keepaspectratio]{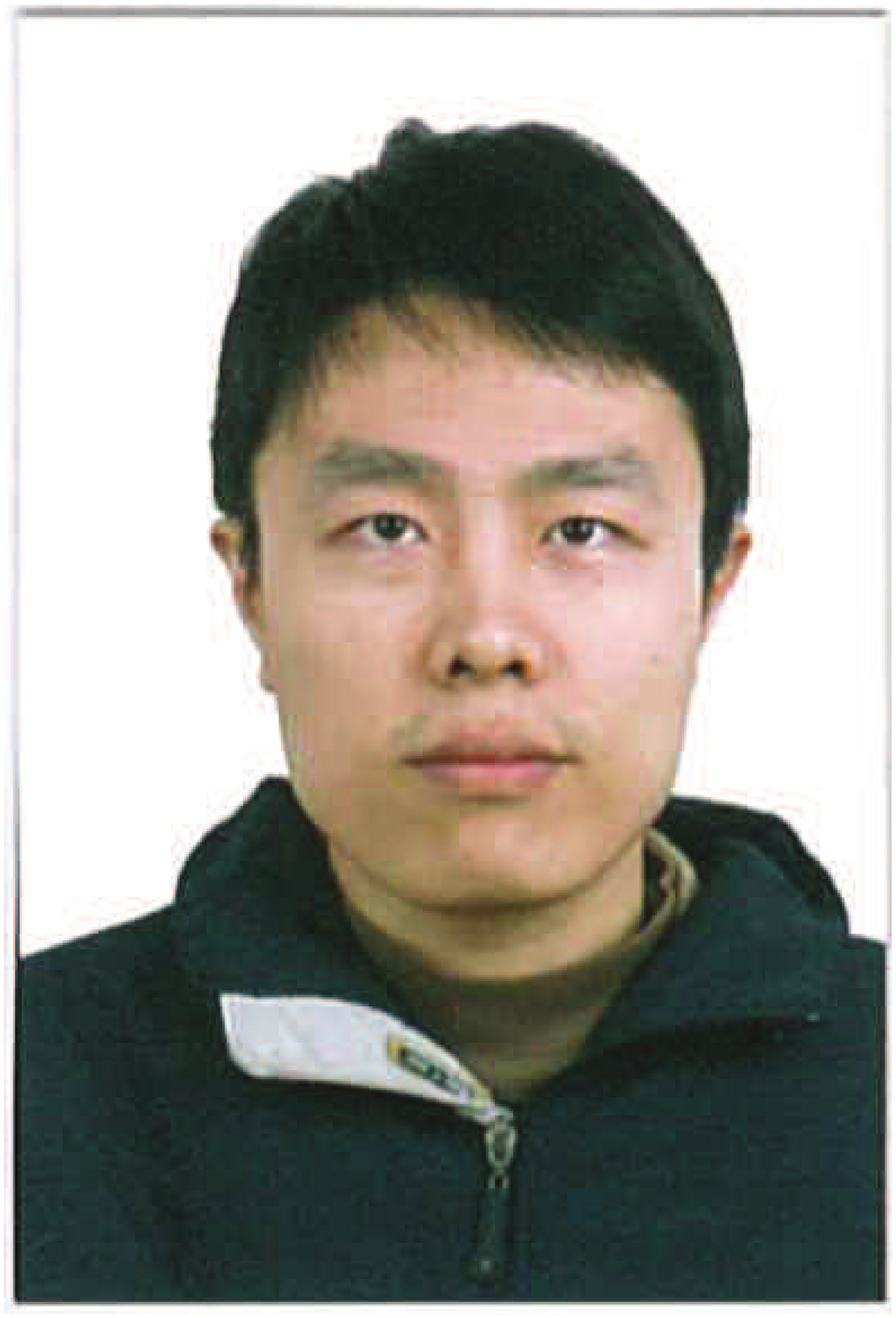}}]{Sheng Jiang} received the Bachelor of Engineering Degree in Microelectronics from Shanghai Jiaotong University, China; Master of Engineering Degree in Electronic Engineering from Hong Kong University of Science and Technology, Hong Kong; and PhD Degree from The Hong Kong Polytechnic University, Hong Kong. He is currently a postdoctoral fellow at The Hong Kong Polytechnic University, Hong Kong.
\end{IEEEbiography}

\begin{IEEEbiography}[{\includegraphics[width=1in,height=1.25in,clip,keepaspectratio]{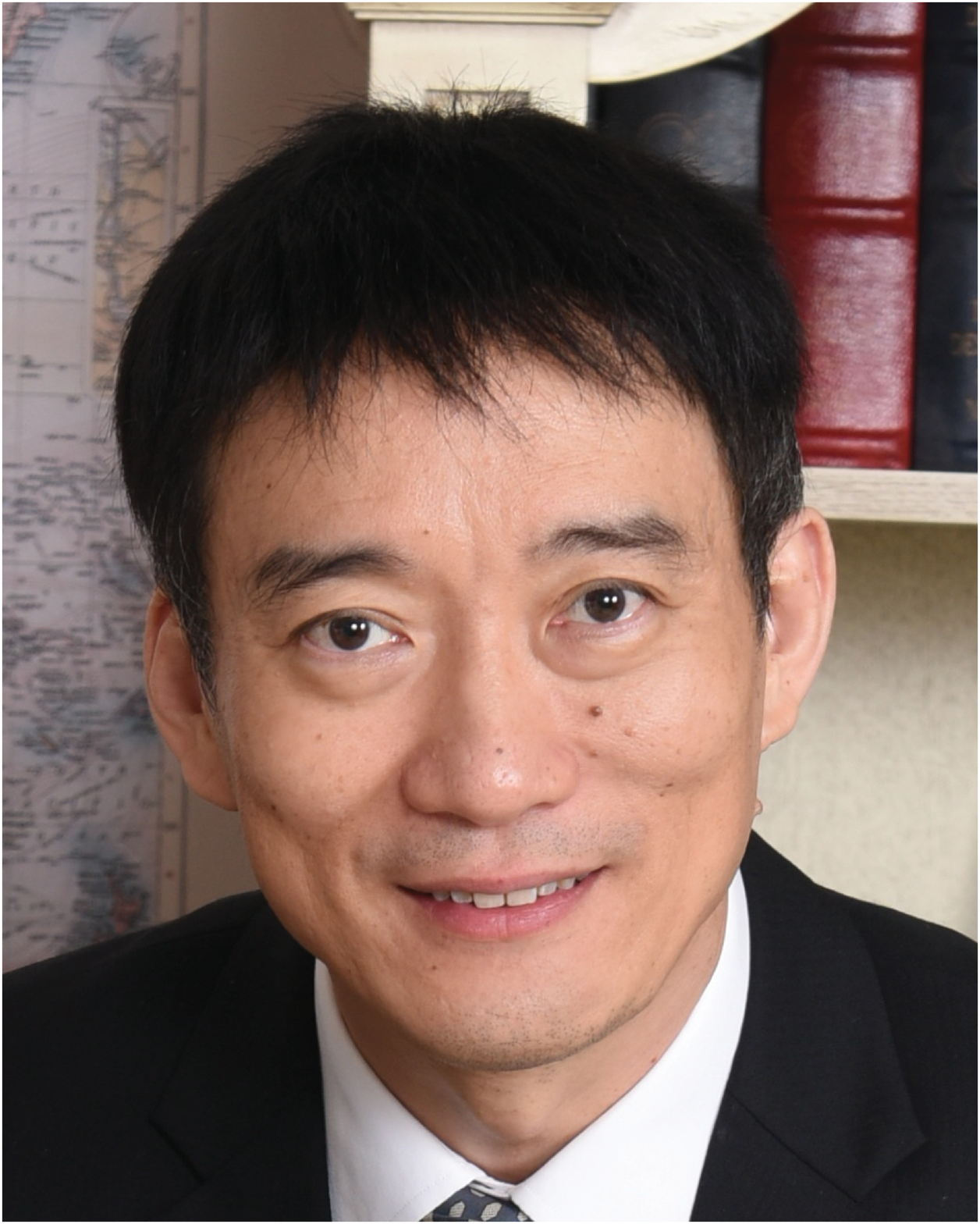}}]{Francis C. M. Lau} received the BEng(Hons) degree in electrical and electronic engineering and the PhD degree from King's College London, University of London, UK. He is a Professor at the Department of Electronic and Information Engineering, The Hong Kong Polytechnic University, Hong Kong. He is also a Fellow of IEEE and a Fellow of IET. 

He is a co-author of two research monographs. He is also a co-holder of six US patents. He has published more than 330 papers. His main research interests include channel coding, cooperative networks, wireless sensor networks, chaos-based digital communications, applications of complex-network theories, and wireless communications. 
He is a co-recipient of one Natural Science Award from the Guangdong Provincial Government, China; eight best/outstanding conference paper awards; one technology transfer award; two young scientist awards from International Union of Radio Science; and one FPGA design competition award.

He was the General Co-chair of International Symposium on Turbo Codes \& Iterative Information Processing (2018) and the Chair of Technical Committee on Nonlinear Circuits and Systems, IEEE Circuits and Systems Society (2012-13). He served as an associate editor for IEEE TRANSACTIONS ON CIRCUITS AND SYSTEMS II (2004-2005 and 2015-2019),  IEEE TRANSACTIONS ON CIRCUITS AND SYSTEMS I (2006-2007), and IEEE CIRCUITS AND SYSTEMS MAGAZINE (2012-2015). He has been a guest associate editor of INTERNATIONAL JOURNAL AND BIFURCATION AND CHAOS since 2010. 
\end{IEEEbiography}

\begin{IEEEbiography}
[{\includegraphics[width=1in,height=1.25in,clip,keepaspectratio]
{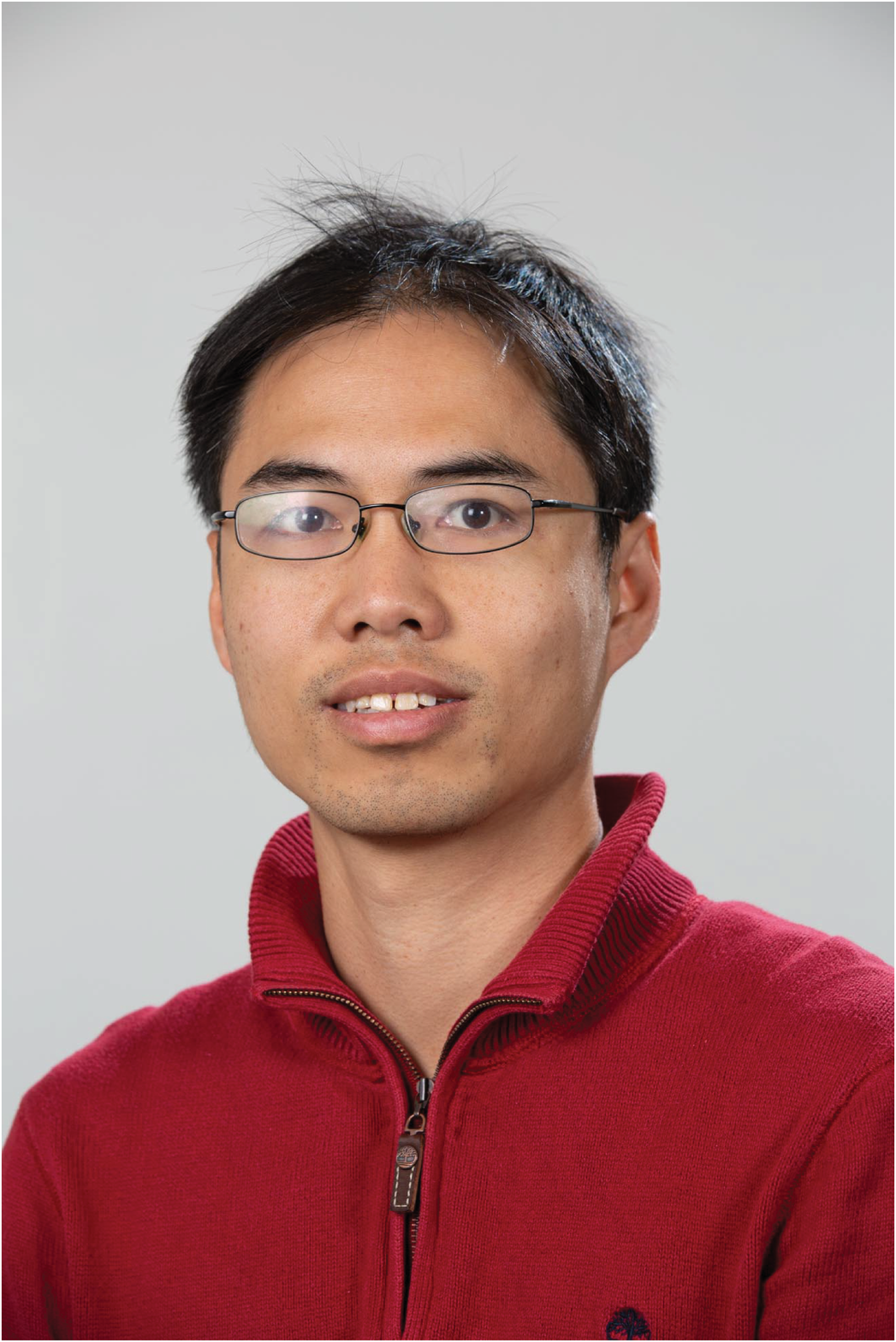}}]
{Chiu-Wing Sham} received his Bachelor degree (Computer Engineering) and  MPhil. degree from The Chinese University of Hong Kong in 2000 and 2002 respectively, and received his Ph.D. degree from the same university in 2006. He has worked as an Electronic Engineer on the FPGA applications of the motion-control system and system security with cryptography in ASM Pacific Technology Ltd (HK). During the years at The Hong Kong Polytechnic University, he engaged in various University projects for the commercialization of technology, in particular, a few optical communication projects which were in collaboration with Huawei. He also worked on the physical design of VLSI design automation. He was invited to work at Synopsys, Inc. (Shanghai) in the summer of 2005 as a Visiting Research Engineer. He is now working at The University of Auckland as a Senior Lecturer. He is also an IEEE Senior Member and an Associate Editor of IEEE TRANSACTIONS ON CIRCUITS AND SYSTEMS II (2017-present).
\end{IEEEbiography}

\end{document}